\documentclass[prd,preprint,showpacs,groupedaddress]{revtex4-1}
\pdfoutput=1
\usepackage{amsmath}
\usepackage{amsfonts}
\usepackage{amssymb}
\usepackage{geometry}
\usepackage{natbib}
\usepackage[english]{babel}
\usepackage{graphicx}
\usepackage{epstopdf}
\usepackage{subfigure}
\usepackage{caption}
\usepackage{multirow}
\usepackage{indentfirst}
\usepackage{mathrsfs}
\usepackage[colorlinks,linkcolor=blue,anchorcolor=blue,citecolor=blue]{hyperref}

\begin{document}

  \setlength{\parindent}{2em}
  \title{Shadow cast by a rotating and nonlinear magnetic-charged black hole in perfect fluid dark matter}
  \author{Tian-Chi Ma} \author{He-Xu Zhang} \author{Peng-Zhang He} \author{Hao-Ran Zhang} \author{Yuan Chen}
  \author{Jian-Bo Deng} \email[Jian-Bo Deng: ]{dengjb@lzu.edu.cn}
  \affiliation{Institute of Theoretical Physics $\&$ Research Center of Gravitation, Lanzhou University, Lanzhou 730000, China}
  \date{\today}

  \begin{abstract}

  We derived an exact solution of the spherically symmetric Hayward black hole surrounded by perfect fluid dark matter (PFDM). By applying the Newman-Janis algorithm, we generalized it to the corresponding rotating black hole. Then, we studied the shadows of rotating Hayward black hole in PFDM. The apparent shape of the shadow depends upon the black hole spin $a$, the magnetic charge $Q$ and the PFDM intensity parameter $k$ ($k<0$). The shadow is a perfect circle in the non-rotating case ($a=0$) and a deformed one in the rotating case ($a\neq{0}$). For a fixed value of $a$, the size of the shadow increases with the increasing $\vert{k}\vert$, but decreases with the increasing $Q$. We further investigated the black hole emission rate. We found that the emission rate decreases with the increasing $\vert{k}\vert$ (or $Q$) and the peak of the emission shifts to lower frequency. Finally, we discussed the observational prospects corresponding to the supermassive black hole $\mathrm{Sgr\ A^{*}}$ at the center of the Milky Way.

  \end{abstract}

  \maketitle

\section{Introduction}

  In recent years, various kinds of astronomical observations strongly reveal that black holes do exist in our universe. So far, the strongest evidence are the recent experimental announcements the detection of gravitational waves (GWs) \cite{abbott2016gw150914} by the LIGO and VIRGO observatories and the captured image of the black hole shadow of a supermassive M87 black hole by the Event Horizon Telescope (EHT) based on the very-long baseline interferometry (VLBI) \cite{event2019first,akiyama2019first}. Among the different methods used to determine the nature of the black hole, observing the shadow of the black hole remains probably the most interesting one. A black hole shadow is the optical appearance that occurs when there is a bright distant light source behind the black hole. To a distant observer, it appears as a two-dimensional dark zone. The observation of the shadow provides a tentative way to find out the parameters of the black hole. Synge \cite{synge1966escape} studied the shadow of the Schwarzschild black hole, he pointed out that the edge of the shadow is rounded. Bardeen \cite{bardeen1973inblack} was the first who studied the shadow of the Kerr black hole. It can be seen that for Kerr black hole, the shadow is no longer circular. Recent works also considered extended Kerr black holes, such as Kerr-de Sitter black holes \cite{grenzebach2014photon,stuchlik2018light,li2020shadow}, deformed black holes \cite{atamurotov2013shadow}, accelerated Kerr black holes \cite{grenzebach2015photon}, Kerr black holes in the presence of extra dimensions \cite{vagnozzi2019hunting,banerjee2020silhouette}. See Refs. \cite{ahmed20205d,chang2020black,contreras2019black,belhaj2020black,gralla2019black,jusufi2020connection,jusufi2020constraining,zaman2020optical,uccanok2020orbits,eiroa2018shadow,maceda2020shadow,abdujabbarov2016shadow,chen2020optical,he2020shadows,cunha2018shadows,cunha2015shadows,cunha2020stationary,dokuchaev2020visible,allahyari2020magnetically,khodadi2020black,badia2020influence,zhang2020optical} for more recent research.

  One of mysterious properties of the black hole is that it itself has a singularity at the origin of the spacetime, at which the curvatures, densities become infinite and the predictive power of physical laws is completely broken down. It is widely believed that the spacetime singularities are the reflection of the incompleteness of General Relativity. Thus, many efforts have been devoted to introduce the black hole without having a spacetime singularity. Surprisingly, Bardeen obtained a black hole solution without a singularity in 1968 \cite{bardeen1968non}. After that, more regular (non-singular) black holes such as Ay\'on-Beato and Garc\'ia black hole \cite{ayon1998regular}, Berej-Matyjasek-Trynieki-Wornowicz black hole \cite{berej2006regular}, and Hayward black hole \cite{hayward2006formation}, were proposed. The spherically symmetric Hayward black hole is described by the metric
  \begin{equation}
  \mathrm{d}s^{2}=-f\left(r\right)\mathrm{d}t^{2}+f\left(r\right)^{-1}\mathrm{d}r^{2}+r^{2}\mathrm{d}\Omega^{2},\quad f\left(r\right)=1-\frac{2Mr^{2}}{r^{3}+Q^{3}},
  \end{equation}
  where $Q$ and $M$ are the magnetic charge and mass, respectively. Hayward black hole behaves like the Schwarzschild black hole at the large distances and at the short distances like the de-Sitter geometry.

  According to the standard model of cosmology, our current universe contains around $68\%$ dark energy, around $28\%$ dark matter and less than $4\%$ baryonic matter \cite{ade2016planck}. Here, dark matter is non-baryonic and non-luminous. Therefore, it is extremely important to study the black hole physics in the presence of dark matter. In recent years, the black hole surrounded by quintessence dark energy have attracted much attention. For example, Kiselev \cite{kiselev2003quintessence} considered the Schwarzschild black hole surrounded by the quintessential energy and then Toshmatov and Stuchl\'ik \cite{toshmatov2017rotating} extended it to the Kerr-like black hole; the Hayward black holes surrounded by quintessence have been studied in Ref. \cite{benavides2020rotating}, etc \cite{ghosh2018lovelock,ghosh2016rotating,chen2008hawking,azreg2013thermodynamical,chakrabarty2019scalar}. On the other hand, as a dark matter candidate, the perfect fluid dark matter has been proposed by Kiselev \cite{kiselev2003quintessential}. In this work, following Refs. \cite{toshmatov2017rotating,benavides2020rotating}, we generalize the Schwarzschild black hole surrounded by PFDM to the spherically symmetric Hayward black hole. In addition, using Newman-Janis algorithm, we obtain the rotating Hayward black hole in PFDM.

\par
  The paper is organized as follows. The next section is the derivation of the spherically symmetric Hayward black hole surrounded by PFDM. In Sec.~\ref{3}, by applying the Newman-Janis algorithm we obtain the rotating Hayward black hole surrounded by PFDM. In Sec.~\ref{4}, we study the photon motion around the rotating Hayward black hole with PFDM.  Black hole shadow of the rotating Hayward black hole with nonvanishing PFDM intensity parameter is considered in Sec.~\ref{5}. In Sec.~\ref{6}, we investigate the energy emission rate of the rotating black hole in PFDM. Conclusions and discussions are presented in Sec.~\ref{7}. Planck units $\hbar=G=c=k_{B}=1$ are used throughout the paper.

\section{Static and spherically symmetric nonlinear magnetic-charged black hole in perfect fluid dark matter}\label{II}
  Let us consider Einstein gravity coupled to a nonlinear electromagnetic field in the presence of the perfect fluid dark matter. It is described by the following equations:
  \begin{gather}
  G_{\mu}^{\,\nu}=2\left(\frac{\partial\mathcal{L}\left(F\right)}{\partial F}F_{\mu\lambda}F^{\nu\lambda}-\delta_{\mu}^{\,\nu}\mathcal{L}\right)+8\pi T_{\mu}^{\,\nu}\left(\mathrm{PFDM}\right),\label{eq:EM1}\\
  \nabla_{\mu}\left(\frac{\partial\mathcal{L}\left(F\right)}{\partial F}F^{\nu\mu}\right)=0.\label{eq:EM2}
  \end{gather}
  Here, $F_{\mu\nu}=2\nabla_{[\mu}A_{\nu]}$ and $\mathcal{L}$ is a function of $F\equiv\frac{1}{4}F_{\mu\nu}F^{\mu\nu}$ given by \cite{nam2018non}
  \begin{equation}
  \mathcal{L}\left(F\right)=\frac{3M}{\vert Q\vert^{3}}\frac{\left(2Q^{2}F \right)^{\frac{3}{2}} }{\left(1+ \left(2Q^{2}F \right)^{\frac{3}{4}}   \right)^{2} },
  \end{equation}
  where $Q$ and $M$ are the parameters associated with magnetic charge and mass, respectively.

\par
  In this work, we consider the black holes surrounded by the perfect fluid dark matter, following Kiselev \cite{kiselev2003quintessence,kiselev2003quintessential} and Li and Yang \cite{li2012galactic}, the energy density of PFDM is given by
  \begin{equation}
  T_{t}^{\,t}=T_{r}^{\,r}=\frac{1}{8\pi}\frac{k}{r^{3}},
  \end{equation}
  with $k$ denoting the intensity of the PFDM. The value of $k$ can be both positive and negative \cite{xu2018kerr}. Here we only consider the theoretical negative values of $k$. The case of positive $k$ can be studied in a similar way as presented in this work.
\par
  To obtain a metric satisfies Eqs.~(\ref{eq:EM1}) and (\ref{eq:EM2}),  let us commence with a spherically symmetric spacetime
  \begin{equation}\label{eq:spherically metric}
   \mathrm{d}s^{2}=-f\left(r\right)\mathrm{d}t^{2}+f\left(r\right)^{-1}\mathrm{d}r^{2}+r^{2}\mathrm{d}\Omega^{2},\quad f\left(r\right)=1-\frac{2m\left(r\right)}{r},
  \end{equation}
  and use the ansatz for Maxwell field \cite{ayon2000bardeen}
  \begin{equation}\label{eq:Maxwell field}
  	F_{\mu\nu}=\left(\delta_{\mu}^{\theta}\delta_{\nu}^{\varphi}-\delta_{\nu}^{\theta}\delta_{\mu}^{\varphi}\right)B\left(r,\theta\right).
  \end{equation}
  With these choices, Eqs. (\ref{eq:EM2}) are easily integrated,
  \begin{equation}
   F_{\mu\nu}=\left(\delta_{\mu}^{\theta}\delta_{\nu}^{\varphi}-\delta_{\nu}^{\theta}\delta_{\mu}^{\varphi}\right)c\left(r\right)\sin{\theta}.
  \end{equation}
  Using the Bianchi identities
  \begin{equation}
  0=\text{\boldmath$dF$}=c^{\prime }(r)\sin (\theta )\text{\boldmath$dr$}%
  \wedge \text{\boldmath$d\theta $}\wedge \text{\boldmath$d\varphi $},
  \label{eq:dF}
  \end{equation}
  one can find that $c(r)=\text{const.}=Q$, where the integration constant has been chosen as $Q$. Further, one can get $F=Q^{2}/2r^{4}$. In order to give a direct physical interpretation to $Q$, we consider the following integral
  \begin{equation}\label{eq:charge}
  \frac{1}{4\pi}\int_{S^{\infty}_{2}}\textbf{\textit{F}}=\frac{Q}{4\pi}\int_{0}^{\pi}\int_{0}^{2\pi}\sin{\theta}\mathrm{d}\theta\mathrm{d}\varphi=Q,
  \end{equation}
  where $S^{\infty}_{2}$ is a two-sphere at the infinity. From Eq. (\ref{eq:charge}), one can confirm that $Q$ is the magnetic monopole charge. For simplicity, with out loss of generality, we consider $Q>0$.

\par
  Now, with the help of the above equations, the time component of Eq.~(\ref{eq:EM1}) reduces to
  \begin{equation}\label{eq:integral}
  -\frac{2}{r^{2}}\frac{\mathrm{d}m\left(r\right)}{\mathrm{d}r}=-\frac{6MQ^{3}}{\left(r^{3}+Q^{3}\right)^{2}}+\frac{k}{r^{3}}.
  \end{equation}
  Integrating Eq.~(\ref{eq:integral}) from $r$ to $\infty$ and using that $M=\lim_{r\to\infty}\left(m\left(r\right)+\frac{k}{2}\ln{\frac{r}{\vert{k}\vert}}\right)$, one finally gets
  \begin{equation}\label{eq:f}
  f\left(r\right)=1-\frac{2Mr^{2}}{r^{3}+Q^{3}}+\frac{k}{r}\ln{\frac{r}{\vert{k}\vert}}.
  \end{equation}

  Thus the metric of exact spherically symmetric solutions for the Einstein equations describing the nonlinear magnetic-charged black holes surrounded by perfect fluid dark matter is given by
  \begin{equation}
  \mathrm{d}s^{2}=-\left(1-\frac{2Mr^{2}}{r^{3}+Q^{3}}+\frac{k}{r}\ln{\frac{r}{\vert{k}\vert}} \right) \mathrm{d}t^{2}+\frac{\mathrm{d}r^{2}}{1-\frac{2Mr^{2}}{r^{3}+Q^{3}}+\frac{k}{r}\ln{\frac{r}{\vert{k}\vert}}}+r^{2}\mathrm{d}\Omega^{2}.
  \end{equation}
  In the absence of PFDM, i.e. $k=0$, we can obtain the non-linear magnetic-charged black hole in the flat background or the Hayward-like black hole \cite{hayward2006formation}.

\section{Rotating and nonlinear magnetic-charged black hole in perfect fluid dark matter}\label{3}
  Many methods have been developed in theory to compute rotating solutions from static ones and the most widely known method is the Newman-Janis algorithm (NJA) and its generalizations. The NJA was first proposed by Newman and Janis in 1965 \cite{newman1965note} and widely used in many articles \cite{toshmatov2017rotating,benavides2020rotating,toshmatov2017generic,kumar2018rotating,xu2017kerr,xu2020black,shaikh2019black,jusufi2020rotating,zhang2020bardeen}. In this work, we will use the NJA modified by Azreg-A\"inou \cite{azreg2014generating} to obtain the Rotating nonlinear magnetic-charged black hole surrounded by perfect fluid dark matter.
  \par
  Consider the general static and spherically symmetric metric:
  \begin{equation}\label{eq:general sphere}
  \mathrm{d}s^{2}=-f\left(r\right)\mathrm{d}t^{2}+g\left(r\right)^{-1}\mathrm{d}r^{2}+h\left(r\right)\mathrm{d}\Omega^{2},\quad \mathrm{d}\Omega^{2}=\mathrm{d}\theta^{2}+\sin^{2}{\theta}\mathrm{d}\varphi^{2}.
  \end{equation}
  At the first step of NJA, we transform the spherically symmetric space-time metric (\ref{eq:general sphere}) from the Boyer-Lindquist (BL) coordinates ($t, r, \theta, \varphi$) to the Eddington-Finkelstein (EF) coordinates ($u, r, \theta, \varphi$). After introducing the coordinate transformation defined by
  \begin{equation}
  \mathrm{d}u=\mathrm{d}t-\frac{\mathrm{d}r}{\sqrt{fg}},
  \end{equation}
  the line element (\ref{eq:general sphere}) takes the form
  \begin{equation}\label{eq:EFC}
  \mathrm{d}s^{2}=-f\left(r\right)\mathrm{d}u^{2}-2\sqrt{\frac{f\left(r\right)}{g\left(r\right)}}\mathrm{d}u\mathrm{d}r+h\left(r\right)\left(\mathrm{d}\theta^{2}+\sin^{2}{\theta}\mathrm{d}\varphi^{2}\right).
  \end{equation}
  In terms of the null tetrads satisfy the relations $l_{\mu}l^{\mu}=n_{\mu}n^{\mu}=m_{\mu}m^{\mu}=l_{\mu}m^{\mu}=n_{\mu}m^{\mu}=0,\ l_{\mu}n^{\mu}=-m_{\mu}\bar{m}^{\mu}=1$, the nonzero components of the inverse metric associated with the line element (\ref{eq:EFC}) can be expressed as
  \begin{equation}\label{eq:covmetric}
  g^{\mu\nu}=-l^{\mu}n^{\nu}-l^{\nu}n^{\mu}+m^{\mu}\bar{m}^{\nu}+m^{\nu}\bar{m}^{\mu},
  \end{equation}
  where
  \begin{equation}\label{eq:nulltetrads}
  \begin{split}
  l^{\mu}&=\delta^{\mu}_{r},\\
  n^{\mu}&=\sqrt{\frac{g\left(r\right)}{f\left(r\right)}}\delta^{\mu}_{u}-\frac{f\left(r\right)}{2}\delta^{\mu}_{r},\\
  m^{\mu}&=\frac{1}{\sqrt{2h\left(r\right)}}\delta^{\mu}_{\theta}+\frac{i}{\sqrt{2h\left(r\right)}\sin{\theta}}\delta^{\mu}_{\varphi},\\
  \bar{m}^{\mu}&=\frac{1}{\sqrt{2h\left(r\right)}}\delta^{\mu}_{\theta}-\frac{i}{\sqrt{2h\left(r\right)}\sin{\theta}}\delta^{\mu}_{\varphi}.
  \end{split}
  \end{equation}
  \par
  Next, we take the critical step of the NJA, which is to perform complex coordinate transformations in the $u-r$ plane
  \begin{equation}\label{eq:transform}
  \begin{gathered}
  u\rightarrow u-ia\cos{\theta},\\
  r\rightarrow r+ia\cos{\theta},
  \end{gathered}
  \end{equation}
  by which $\delta_{\nu}^{\mu}$ transform as vectors:
  \begin{equation}
  \begin{split}
  \delta^{\mu}_{r}&\rightarrow\delta^{\mu}_{r},{\quad}\delta^{\mu}_{u}\rightarrow\delta^{\mu}_{u},{\quad}\\
  \delta^{\mu}_{\theta}&\rightarrow\delta^{\mu}_{\theta}+ia\sin{\theta}\left(\delta^{\mu}_{u}-\delta^{\mu}_{r} \right),{\quad}\delta^{\mu}_{\varphi}\rightarrow\delta^{\mu}_{\varphi}.
  \end{split}
  \end{equation}
  At the same time, we assume that the functions $f\left(r \right) $, $g\left(r \right) $ and $h\left(r \right)$ in Eq. (\ref{eq:nulltetrads})  also turn into a new form: $f\left(r\right)\rightarrow F\left(r, a, \theta\right)$, $g\left(r\right)\rightarrow G\left(r, a, \theta\right)$, and $h\left(r\right)\rightarrow\Sigma=r^{2}+a^{2}\cos^{2}{\theta}$ \cite{azreg2014generating}. Thus, the effect of the transformation~(\ref{eq:transform}) on Eq.~(\ref{eq:nulltetrads}) is
  \begin{equation}
  \begin{split}
  l^{\mu}&=\delta^{\mu}_{r},\quad n^{\mu}=\sqrt{\frac{G}{F}}\delta^{\mu}_{u}-\frac{1}{2}F\delta^{\mu}_{r},\\
  m^{\mu}&=\frac{1}{\sqrt{2\Sigma}}\left(\delta^{\mu}_{\theta}+ia\sin{\theta}\left(\delta^{\mu}_{u}-\delta^{\mu}_{r}\right)+\frac{i}{\sin{\theta}}\delta^{\mu}_{\varphi}\right),\\
  \bar{m}^{\mu}&=\frac{1}{\sqrt{2\Sigma}}\left(\delta^{\mu}_{\theta}-ia\sin{\theta}\left(\delta^{\mu}_{u}-\delta^{\mu}_{r}\right)-\frac{i}{\sin{\theta}}\delta^{\mu}_{\varphi}\right).
  \end{split}
  \end{equation}
  Then by means of Eq. (\ref{eq:covmetric}), we obtain the spacetime metric tensor $g^{\mu\nu}$ as
  \begin{equation}
  \begin{split}
  g^{uu}&=\frac{a^{2}\sin^{2}{\theta}}{\Sigma},\quad g^{rr}=G+\frac{a^{2}\sin^{2}{\theta}}{\Sigma},\\
  g^{\theta\theta}&=\frac{1}{\Sigma},\quad g^{\varphi\varphi}=\frac{1}{\Sigma\sin^{2}{\theta}},\\
  g^{ur}&=g^{ru}=-\sqrt{\frac{G}{F}}-\frac{a^{2}\sin^{2}{\theta}}{\Sigma},\\
  g^{u\varphi}&=g^{\varphi u}=\frac{a}{\Sigma},\quad g^{r\varphi}=g^{\varphi r}=-\frac{a}{\Sigma}.
  \end{split}
  \end{equation}
  Accordingly, the covariant metric in the EF coordinates ($u, r, \theta, \phi$) reads
  \begin{equation}\label{eq:metricinEDC}
  \begin{split}
  \mathrm{d}s^{2}=&-F\mathrm{d}u^{2}-2\sqrt{\frac{F}{G}}\mathrm{d}u\mathrm{d}r+2a\left(F-\sqrt{\frac{F}{G}} \right)\sin^{2}\theta\mathrm{d}u\mathrm{d}\varphi+\Sigma\mathrm{d}\theta^{2}\\
  &+2a\sin^{2}\theta\sqrt{\frac{F}{G}}\mathrm{d}r\mathrm{d}\varphi+\sin^{2}\theta\left[\Sigma+a^{2}\left(2\sqrt{\frac{F}{G}}-F \right)\sin^{2} \right]\mathrm{d}\varphi^{2}.
  \end{split}
  \end{equation}

  \par
  The final step of NJA is to bring (\ref{eq:metricinEDC}) to the BL coordinates by a coordinate transformations:
  \begin{equation}
  \mathrm{d}u=\mathrm{d}t+\lambda\left(r\right)\mathrm{d}r,\quad \mathrm{d}\varphi=\mathrm{d}\phi+\chi\left(r\right)\mathrm{d}r,
  \end{equation}
  where the functions $\lambda\left(r\right)$ and $\chi\left(r\right)$ can be found using the requirement that all nondiagonal components of the metric tensor (except for the coefficient $g_{t\phi}$ ($g_{\phi{t}}$) ) are equal to zero, Thus
  \begin{equation}
  \begin{gathered}
  \lambda\left(r\right)=-\frac{p\left(r \right) +a^{2}}{g\left(r\right)h\left(r \right) +a^{2}},\quad \chi\left(r\right)=-\frac{a}{g\left(r\right)h\left(r \right) +a^{2}},\\
  \end{gathered}
  \end{equation}
  with
  \begin{equation}
  p\left(r \right)=\sqrt{\frac{g\left(r\right) }{f\left( r\right) }}h\left(r \right)
  \end{equation}
  and
  \begin{equation}
  F\left(r,a,\theta \right)=\frac{\left(gh+a^{2}\cos^{2}\theta \right)\Sigma }{\left(p^{2}+a^{2}\cos^{2}\theta \right)^{2} },\quad{G\left(r,a,\theta \right) }=\frac{gh+a^{2}\cos^{2}\theta}{\Sigma}.
  \end{equation}
  Finally, the rotating solution corresponding to the spherically symmetric metric (\ref{eq:general sphere}) can therefore be obtained as
  \begin{equation}\label{eq:rotating solution}
  \begin{split}
  \mathrm{d}s^{2}=&-\frac{\left(gh+a^{2}\cos^{2}{\theta}\right)\Sigma}{\left( p+a^{2}\cos^{2}{\theta}\right)^{2} }\mathrm{d}t^{2}+\frac{\Sigma}{gh+a^{2}}\mathrm{d}r^{2}-2a\sin^{2}{\theta}\left( \frac{p-gh}{\left(p+a^{2}\cos^{2}{\theta} \right)^{2} }\right) {\Sigma}\mathrm{d}{\phi}\mathrm{d}{t}\\
  &+\Sigma\mathrm{d}\theta^{2}+\Sigma\sin^{2}{\theta}\left(1+a^2\sin^{2}{\theta}\frac{2p-gh+a^{2}\cos^{2}{\theta}}{\left( p+a^{2}\cos^{2}{\theta}\right)^{2} } \right)\mathrm{d}{\phi}^{2}.
  \end{split}
  \end{equation}

  In the case of Hayward black holes in PFDM, comparing the line elements (\ref{eq:spherically metric}) with (\ref{eq:general sphere}), one can find
  \begin{equation}
  \begin{split}
  g\left(r \right)=f\left(r \right),\quad{p\left(r \right) }=h\left(r \right)=r^{2}.
  \end{split}
  \end{equation}
  Substituting the above expressions into (\ref{eq:rotating solution}), we obtain the metric of rotating Hayward black holes in perfect fluid dark matter in the form
  \begin{equation}\label{eq:rotating metric}
  \begin{split}
  \mathrm{d}s^{2}=&-\left(1-\frac{r^{2}-f\left(r \right)r^{2} }{\Sigma}\right)\mathrm{d}t^{2}+\frac{\Sigma}{\Delta_{r}}\mathrm{d}r^{2}-\frac{2a\sin^{2}\theta\left(r^{2}-f\left(r \right)r^{2}  \right) }{\Sigma}\mathrm{d}t\mathrm{d}\phi\\
  &+\Sigma\mathrm{d}\theta^{2}+\frac{\sin^{2}\theta}{\Sigma}\left(\left(r^{2}+a^{2} \right)^{2}-a^{2}{\Delta_{r}}\sin^{2}\theta  \right) \mathrm{d}\phi^{2},
  \end{split}
  \end{equation}
  with
  \begin{equation}\label{eq:line element terms}
  \begin{gathered}
  \Sigma=r^{2}+a^{2}\cos^{2}{\theta},\\
  \Delta_{r}=r^{2}f\left(r \right)+a^{2},\\
  f\left(r\right)=1-\frac{2Mr^{2}}{r^{3}+Q^{3}}+\frac{k}{r}\ln{\frac{r}{\vert{k}\vert}}.
  \end{gathered}
  \end{equation}

\section{Photon orbits}\label{4}

\par
  For obtaining the geodesic equations, we use the Hamilton-Jacobi equation and Carter constant separable method \cite{carter1968global}. The Hamilton-Jacobi equation takes the form as
  \begin{equation}\label{eq:hamilton}
  \frac{\partial S}{\partial \tau}=-\frac{1}{2} g^{\mu\nu}\frac{\partial{S}}{\partial{x^{\mu}}}\frac{\partial{S}}{\partial{x^{\nu}}},
  \end{equation}
  where $S$ is the Jacobi action and $\tau$ is an affine parameter. We select the corresponding action as

  \begin{equation}\label{eq:action}
  S=\frac{1}{2} m^{2}\tau -Et + L\phi + S_{r}(r) + S_{\theta}(\theta),
  \end{equation}
  where $m$ is proportional to the rest mass of the particle. Energy $E$ and angular momentum $L$ are constant of motion related to the associated Killing vectors $\partial{}/\partial{t}$ and $\partial{}/\partial{\phi}$. $S_r\left(r \right) $ and $S_{\theta}\left(\theta \right)$ are functions of $r$ and $\theta$, respectively.
  \par
  For rotating Hayward black hole in PFDM, the Hamilton-Jacobi equation yields:
  \begin{equation}
  \frac{1}{2}g^{tt}\frac{\partial{S}}{\partial{t}}\frac{\partial{S}}{\partial{t}}
  +\frac{1}{2}g^{rr}\frac{\partial{S}}{\partial{r}}\frac{\partial{S}}{\partial{r}}
  +\frac{1}{2}g^{\theta\theta}\frac{\partial{S}}{\partial{\theta}}\frac{\partial{S}}{\partial{\theta}}
  +\frac{1}{2}g^{\phi\phi}\frac{\partial{S}}{\partial{\phi}}\frac{\partial{S}}{\partial{\phi}}
  +g^{\phi{t}}\frac{\partial{S}}{\partial{\phi}}\frac{\partial{S}}{\partial{t}}
  =-\frac{\partial{S}}{\partial{\tau}}.
  \end{equation}
  Then, we solve it for $S_{r}\left(r \right) $ and $S_{\theta}\left(\theta \right) $ as follows \cite{chandrasekhar1998mathematical}:
  \begin{equation}
  \begin{split}
  \frac{\mathrm{d}S_{r}(r)}{\mathrm{d}r}=\frac{\sqrt{R(r)}}{\Delta_{r}},\label{eq:R}\\
  \frac{\mathrm{d}S_{\theta}(\theta)}{\mathrm{d}\theta}=\sqrt{\Theta (\theta)},
  \end{split}
  \end{equation}
  where
  \begin{gather}
  R(r)=(aL-(a^{2}+r^{2})E)^{2}-(K+\left(aE-L \right)^{2} +m^{2}r^{2})\Delta_{r},\label{eq:Rm}\\
  \Theta(\theta)=K-a^{2}m^{2}\cos^{2}(\theta)-\left(\frac{L^{2}}{\sin^{2}\theta}-a^{2}E^{2} \right)\cos^{2}\theta .\label{eq:Thetam}
  \end{gather}
  with $K$ as a constant of separation.
  We can replace $S_{r}(r)$ and $S_{\theta}(\theta)$ in formula (\ref{eq:action}) with
  \begin{equation}
  \begin{split}
  S_{r}=\int^{r}\frac{\sqrt{R(r)}}{\Delta_{r}}\mathrm{d}r\\
  S_{\theta}=\int^{\theta}\sqrt{\Theta (\theta)}\mathrm{d}\theta.
  \end{split}
  \end{equation}
  By following \cite{hou2018rotating}, we can write the equations of the geodesic motion in the form
  \begin{gather}
  \Sigma\dot{t}=\frac{((r^{2}+a^{2})E-aL)(r^{2}+a^{2})}{\Delta_{r}}-a(aE\sin^{2}(\theta)-L),\label{eq:tdot}\\
  \Sigma\dot{r}=\sqrt{R(r)},\label{eq:rdot}\\
  \Sigma\dot{\theta}=\sqrt{\Theta (\theta)},\label{eq:thetadot}\\
  \Sigma\dot{\phi}=\frac{a((r^{2}+a^{2})E-aL)}{\Delta_{r}}-\frac{aE\sin^{2}(\theta)-L}{\sin^{2}(\theta)},\label{eq:phidot}
  \end{gather}
  where overdot ( $\dot{}$ ) stands for the derivative with respect to the proper time $\tau$.

  As we are interested in black hole shadows, we consider only null geodesics, for which $m=0$. The motion of photon is determined by two impact parameters $\xi=L/E$ and $\eta=K/E^{2}$. We define $R_p\left(r \right)=R\left(r \right)/E^{2}  $ and $\Theta_{p}\left(\theta \right)=\Theta\left(\theta \right)/E^{2}  $. Then, for photon, we rewrite Eqs. (\ref{eq:Rm}) and (\ref{eq:Thetam}) as:
  \begin{gather}
  R_{p}(r)=(a\xi-(a^{2}+r^{2}))^{2}-(\eta+\left(a-\xi \right)^{2})\Delta_{r},\label{eq:Rp}\\
  \Theta_{p}(\theta)=\eta-\left(\frac{\xi^{2}}{\sin^{2}\theta}-a^{2} \right)\cos^{2}\theta .\label{eq:Thetap}
  \end{gather}
  For spherical orbit, a test photon has zero radial velocity ($\dot{r}=0$ ) and zero radial acceleration ($\ddot{r}=0$), By (\ref{eq:rdot}), this requires that
  \begin{equation}\label{eq:photonregion}
  R_{p}(r)=0 , \quad
  \frac{\mathrm{d}R_{p}(r)}{\mathrm{d}r}=0.
  \end{equation}
  Combining Eqs. (\ref{eq:Rp}) and (\ref{eq:photonregion}), we get the expressions of $\xi$ and $\eta$ as
  \begin{equation}\label{eq:xi and eta}
  \begin{split}
  \xi&=\frac{-4r\Delta_{r}+a^{2}\Delta_{r}^{\prime}+r^{2}\Delta_{r}^{\prime} }{a\Delta_{r}^{\prime}},\\
  \eta&=\frac{r^{2}\left(16a^{2}\Delta_{r}-16\Delta_{r}^{2}+8r\Delta_{r}\Delta_{r}^{\prime}-r^{2}\Delta_{r}^{\prime{2}} \right) }{a^{2}\Delta_{r}^{\prime{2}}}
  \end{split}
  \end{equation}
  Inserting (\ref{eq:xi and eta}) into (\ref{eq:Thetap}), using the fact that $\Theta_{p}\left(\theta \right) $ is non negative, one can get an inequality that determines the photon region \cite{grenzebach2014photon}
  \begin{equation}\label{eq:photon region}
  \mathcal{K}:{\;}16r^{2}a^{2}\Delta_{r}\sin^{2}(\theta)-(4r\Delta_{r}-\Sigma\Delta'_{r})^{2}\geq 0.
  \end{equation}
  When $a=0$, Eq. (\ref{eq:photon region}) degenerates into an equality,
  \begin{equation}
  (4r\Delta_{r}-\Sigma\Delta'_{r})^{2}= 0.
  \end{equation}
  This means that the photon regions degenerate into photon spheres.

  A spherical lightlike geodesic at $r=r_p$ is stable with respect to radial perturbations if $\frac{\mathrm{d}^{2}R_p\left(r \right)}{\mathrm{d}r^{2}}\vert_{r=r_p}  <0$, and unstable if $\frac{\mathrm{d}^{2}R_p\left(r \right)}{\mathrm{d}r^{2}}\vert_{r=r_p} >0$. With the help of (\ref{eq:Rp}) and (\ref{eq:xi and eta}), one can get
  \begin{equation}\label{eq:region stable}
  \frac{\mathrm{d}^{2}R_p\left(r \right)}{\mathrm{d}r^{2}}=\frac{8E^{2}\left( 2r\Delta_{r}\Delta_{r}^{\prime}+r^{2}\Delta_{r}^{\prime{2}}-2r^{2}\Delta_{r}\Delta_{r}^{\prime\prime}\right)}{\Delta_{r}^{\prime{2}}},
  \end{equation}
  where ( $^\prime{}$ ) denotes the derivative with respect to $r$.

  \begin{figure}[]
  	\centering
  	\includegraphics[width=.49\textwidth]{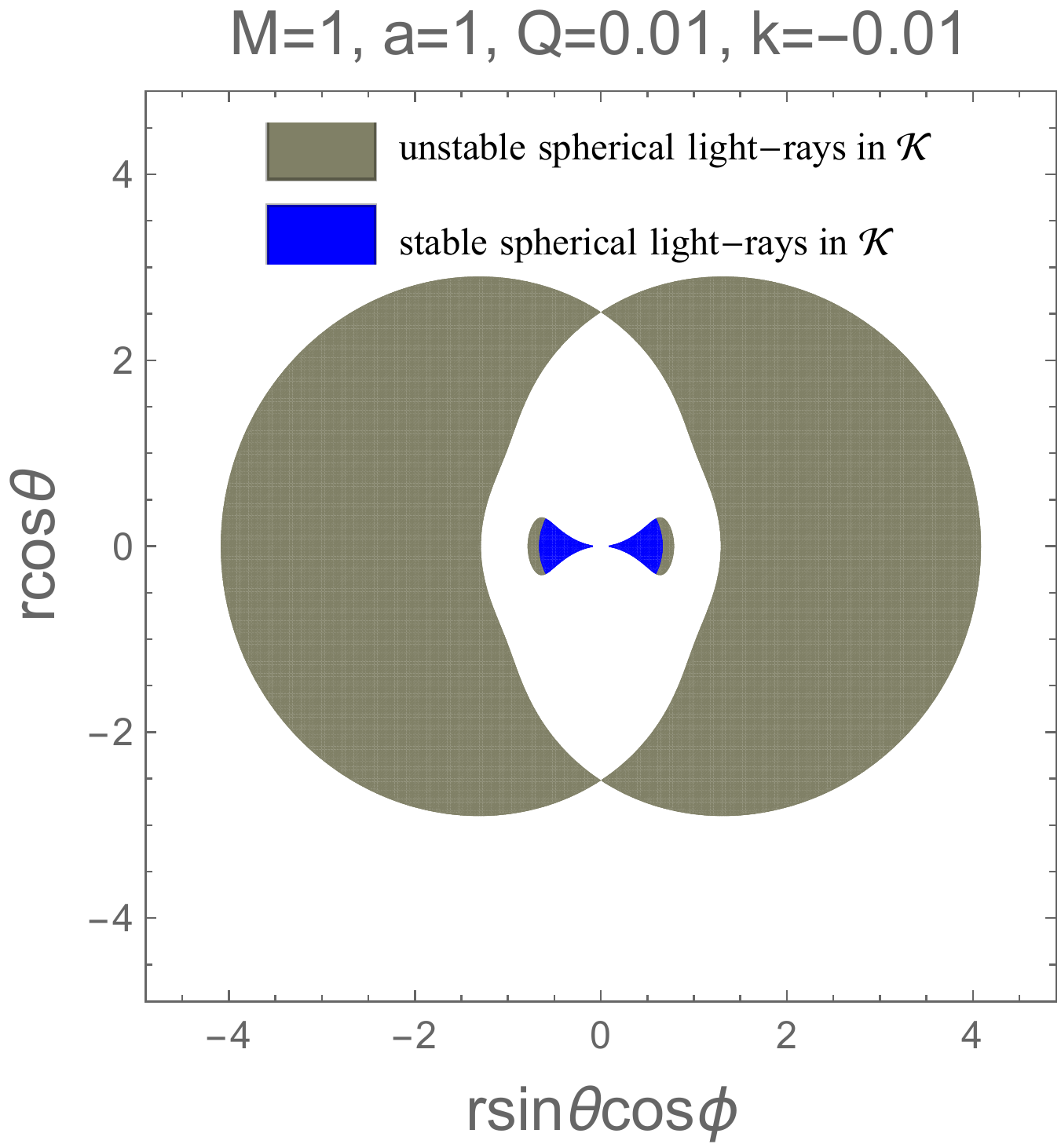}
  	\caption{Photon region of rotating nonlinear magnetic-charged black hole surrounded by PFDM with $a=1$, $Q=0.01$ and $k=-0.01$.}
  	\label{photonregion}
  \end{figure}

  Fig. \ref{photonregion} shows the photon region $\mathcal{K}$ in the $(r,\theta)$ plane with $a=1$, $Q=0.01$ and $k=-0.01$, where stable and unstable spherical light rays are distinguished. The unstable spherical light rays determine the contour of the shadow.

  \section{Black hole shadow}\label{5}

  Now we investigate the shadow of a rotating Hayward black hole in PFDM. Here, we assume that the light sources exist at infinity and are distributed uniformly in all directions. To determine the shape of the black hole shadow, we introduce the celestial coordinates $\alpha$ and $\beta$ as
  \begin{equation}
  \begin{split}
  \alpha&=\lim_{r_{o}\to\infty}\left(-r_{o}^{2}\sin{\theta\frac{\mathrm{d}\phi}{\mathrm{d}r}}\vert_{\theta\rightarrow{i}}\right),\\
  \beta&=\pm\lim_{r_{o}\to\infty}\left(r_{o}^{2}\frac{\mathrm{d}\theta}{\mathrm{d}r}\vert_{\theta\rightarrow{i}} \right) ,
  \end{split}
  \end{equation}
  where $r_o$ is the distance from the black hole to observer, $i$ is the angle between the rotation axis of the black hole and the line of sight of the observer. Here we assume the observer is at infinity.
  \par
  Using the geodesic equations (\ref{eq:tdot})-(\ref{eq:phidot}), the celestial coordinates, as a function of $\xi$ and $\eta$, take the form
  \begin{equation}
  \begin{split}
  \alpha&=-\frac{\xi}{\sin{i}},\\
  \beta&=\pm\sqrt{\eta+a^{2}\cos^{2}{i}-{\xi}^{2}\cot^{2}{i}}.
  \end{split}
  \end{equation}
  In the equatorial plane ($i={\pi}/2$), $\alpha$ and $\beta$ reduce to
  \begin{equation}\label{eq:alpha and beta}
  \begin{split}
  \alpha&=-{\xi},\\
  \beta&=\pm\sqrt{\eta}.
  \end{split}
  \end{equation}
  Furthermore, using Eqs. (\ref{eq:line element terms}), (\ref{eq:xi and eta}) and (\ref{eq:alpha and beta}), we have
  \begin{equation}\label{eq:hj2}
  \begin{split}
  \alpha^{2}+\beta^{2}&=\xi^{2}+\eta\\
  &=2r^{2}+a^{2}+\frac{8\Delta_r[2- \left(rf^{\prime}\left(r \right) +2f\left(r \right)  \right) ] }{\left(rf^{\prime}\left(r \right)+2f\left(r \right)   \right)^{2} },
  \end{split}
  \end{equation}
  where
  \begin{equation}\label{eq:hj2}
  \begin{split}
  f^{\prime}\left(r \right)=\frac{2Mr^{4}-4MQ^{3}r}{\left( Q^{3}+r^{3}\right)^{2} }+\frac{k}{r^{2} }\left(1-\ln\frac{r}{\vert{k}\vert} \right).
  \end{split}
  \end{equation}

  \begin{figure}[]
  	\centering
  	\includegraphics[width=.49\textwidth]{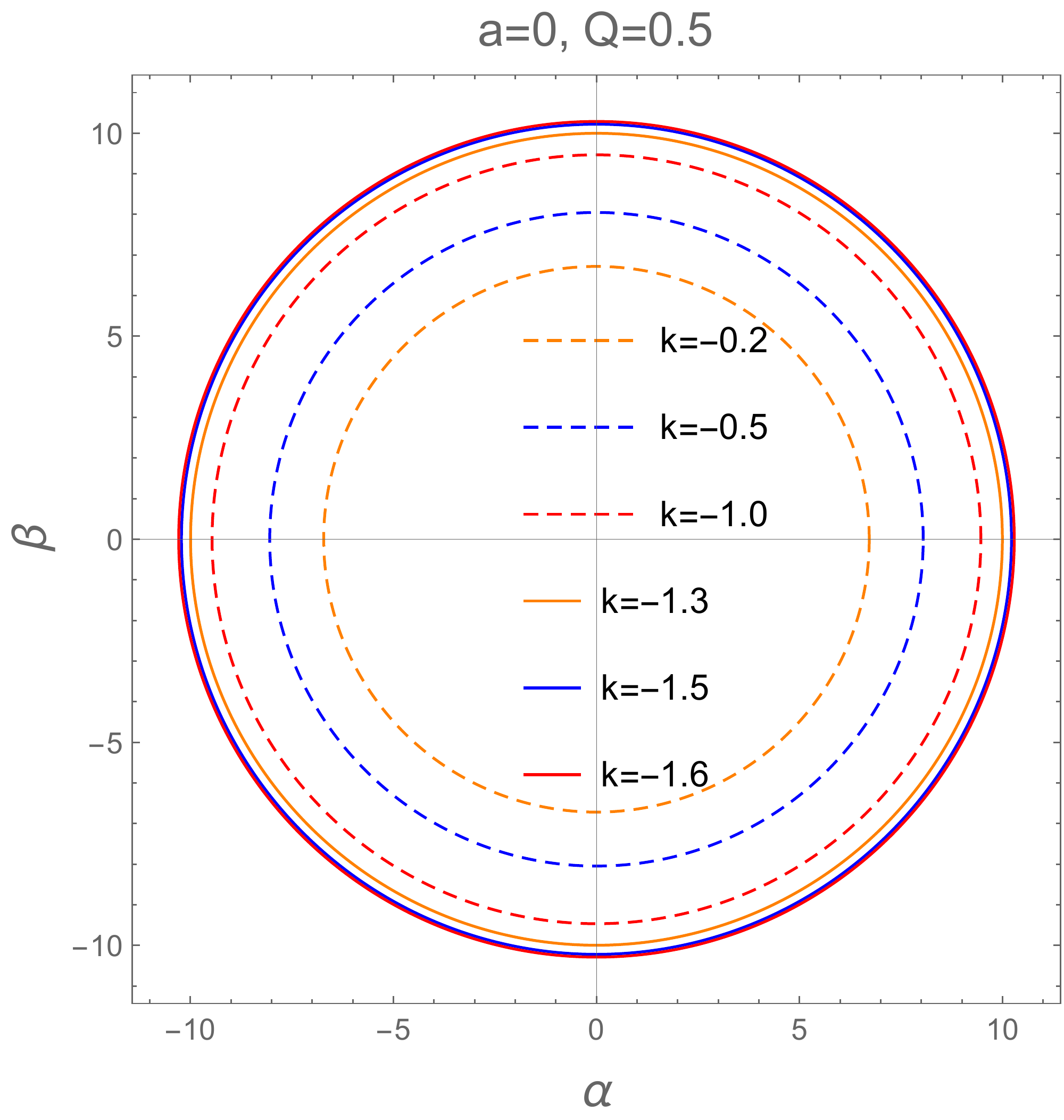}
  	\includegraphics[width=.49\textwidth]{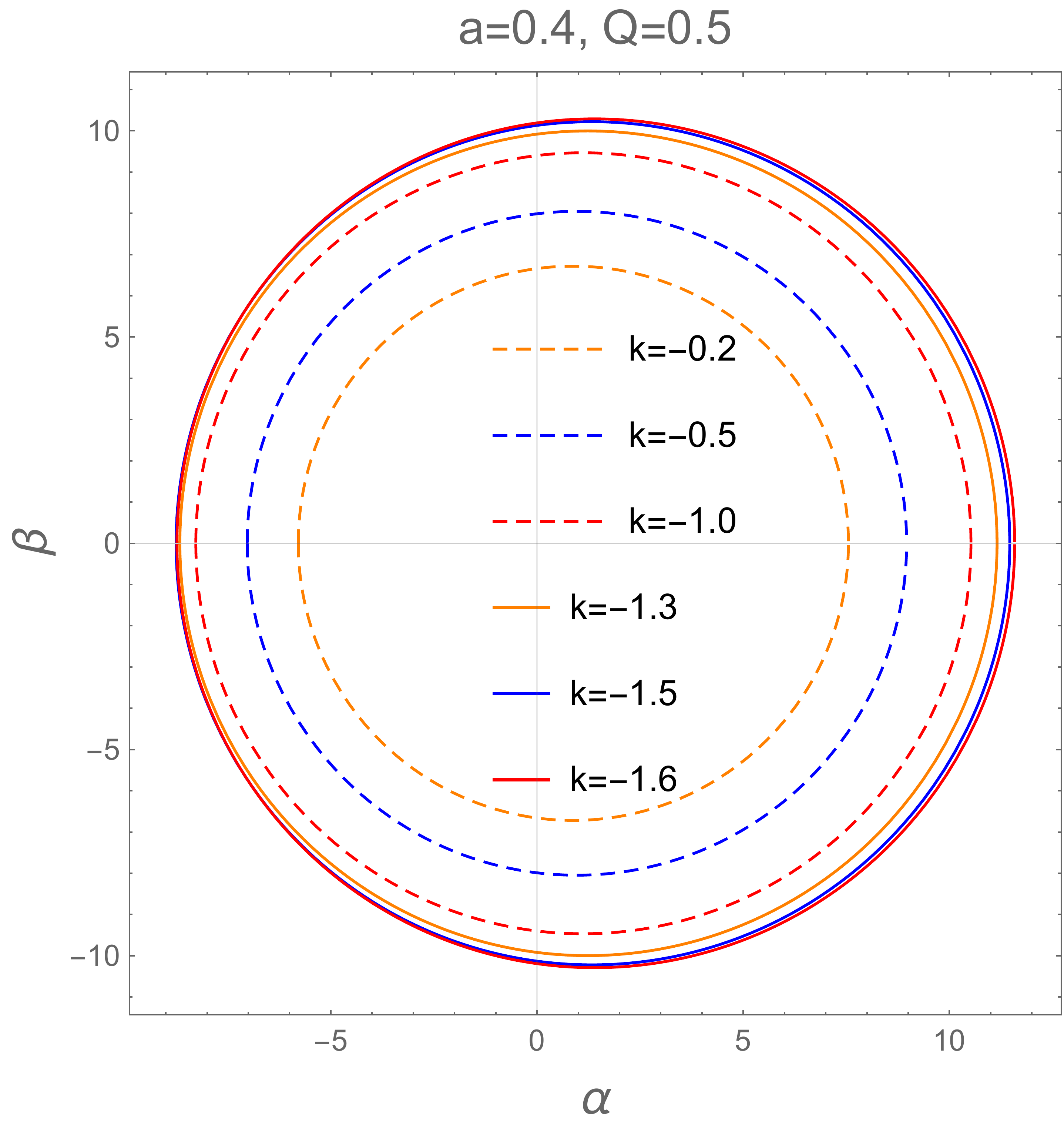}
  	\includegraphics[width=.49\textwidth]{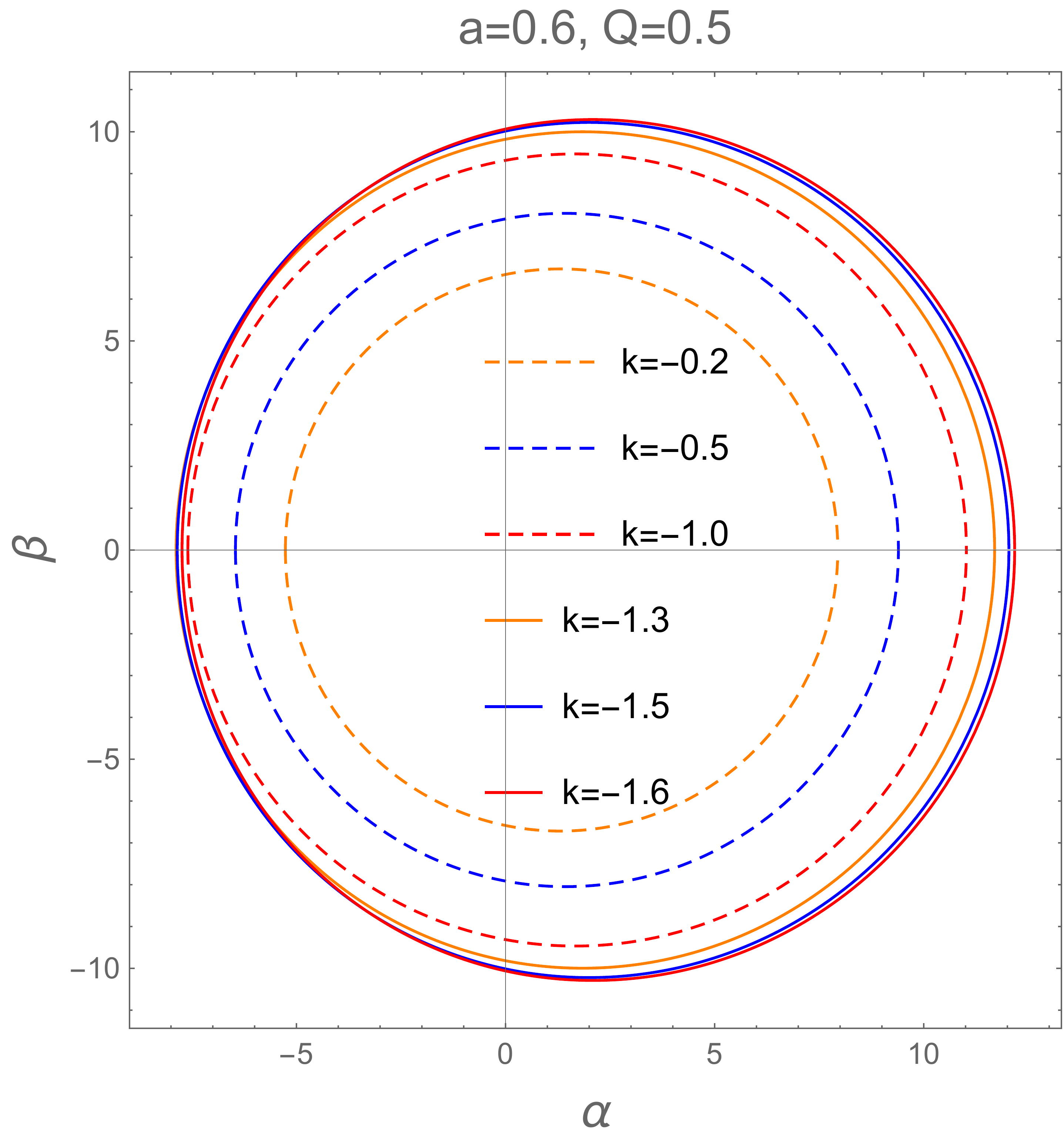}
  	\includegraphics[width=.49\textwidth]{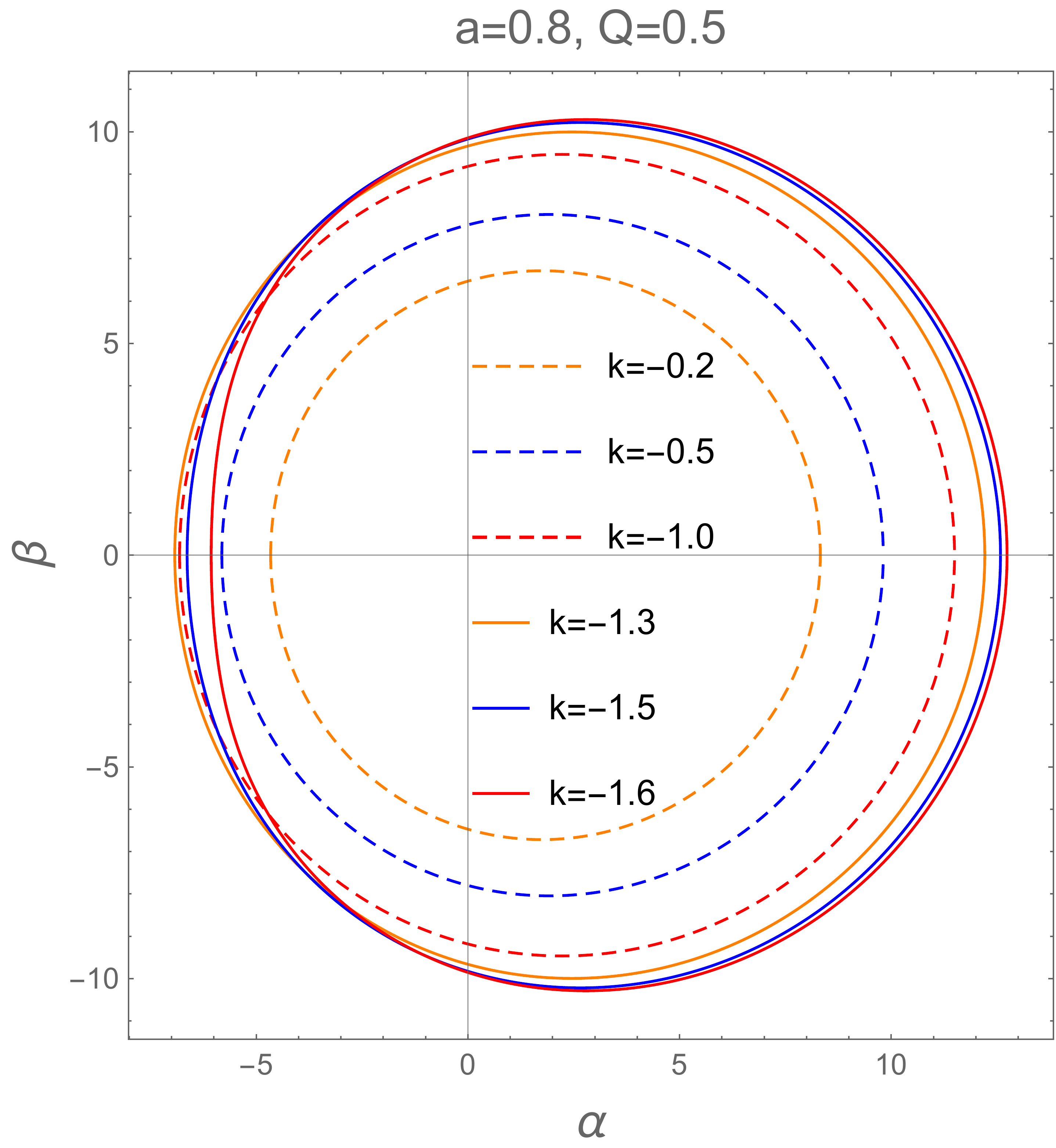}
  	\caption{Examples of shadows of the rotating nonlinear magnetic-charged black hole surrounded by PFDM for different values of the parameters $a$ and $k$ with the magnetic charge $Q=0.5$  .}
  	\label{shadowofk}
  \end{figure}

\begin{figure}[]
  	\centering
  	\includegraphics[width=.49\textwidth]{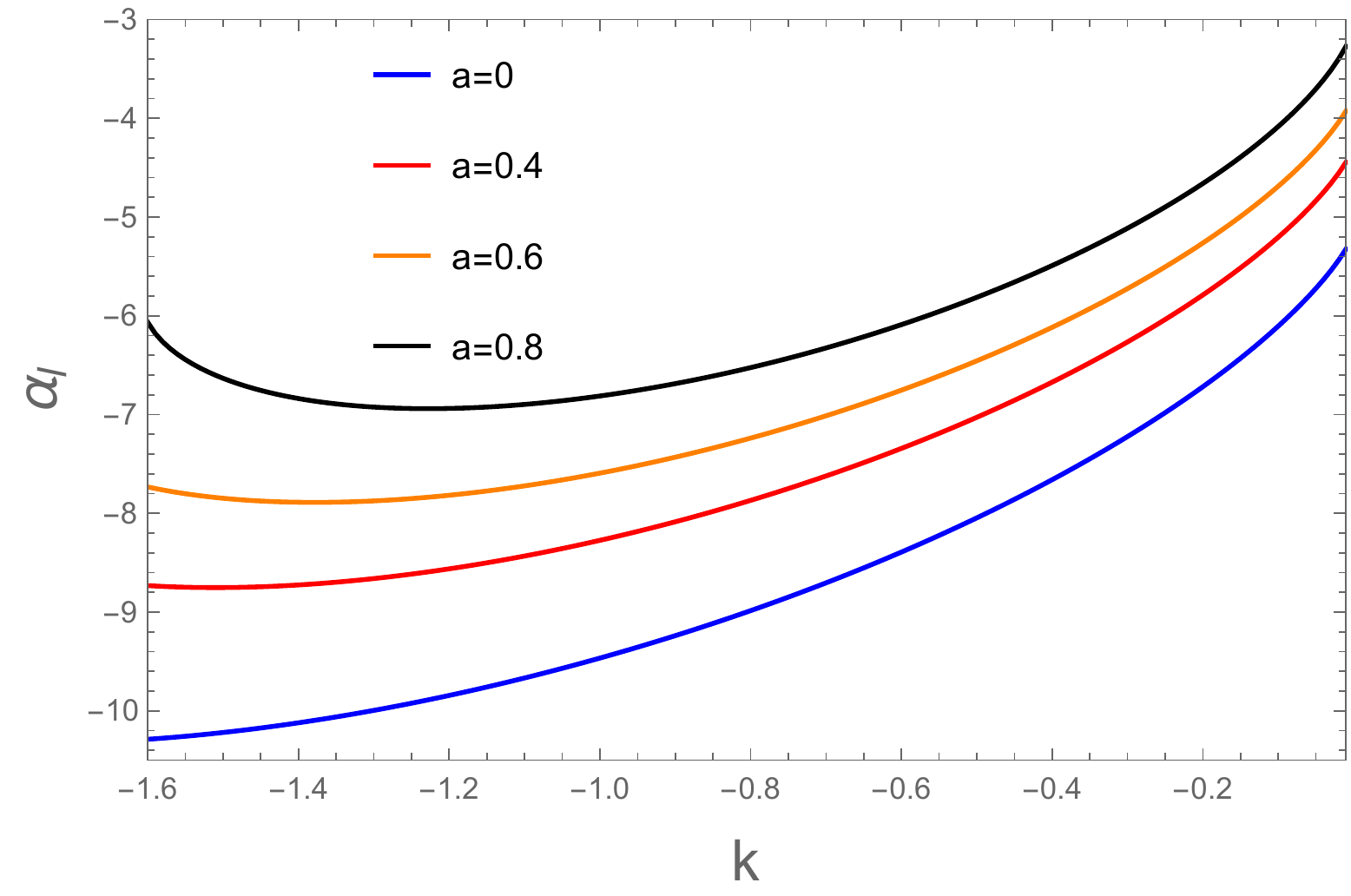}
  	\caption{Dependence of the horizontal coordinate $\left( \alpha_l\right)$  of the most left position (marked by $L$ in figure \ref{lRspic} ) of the shadow on the spin parameter $a$ and the PFDM intensity parameter $k$ with $Q=0.5$ .}
  	\label{most left}
  \end{figure}

\begin{figure}[]
  	\centering
  	\includegraphics[width=.49\textwidth]{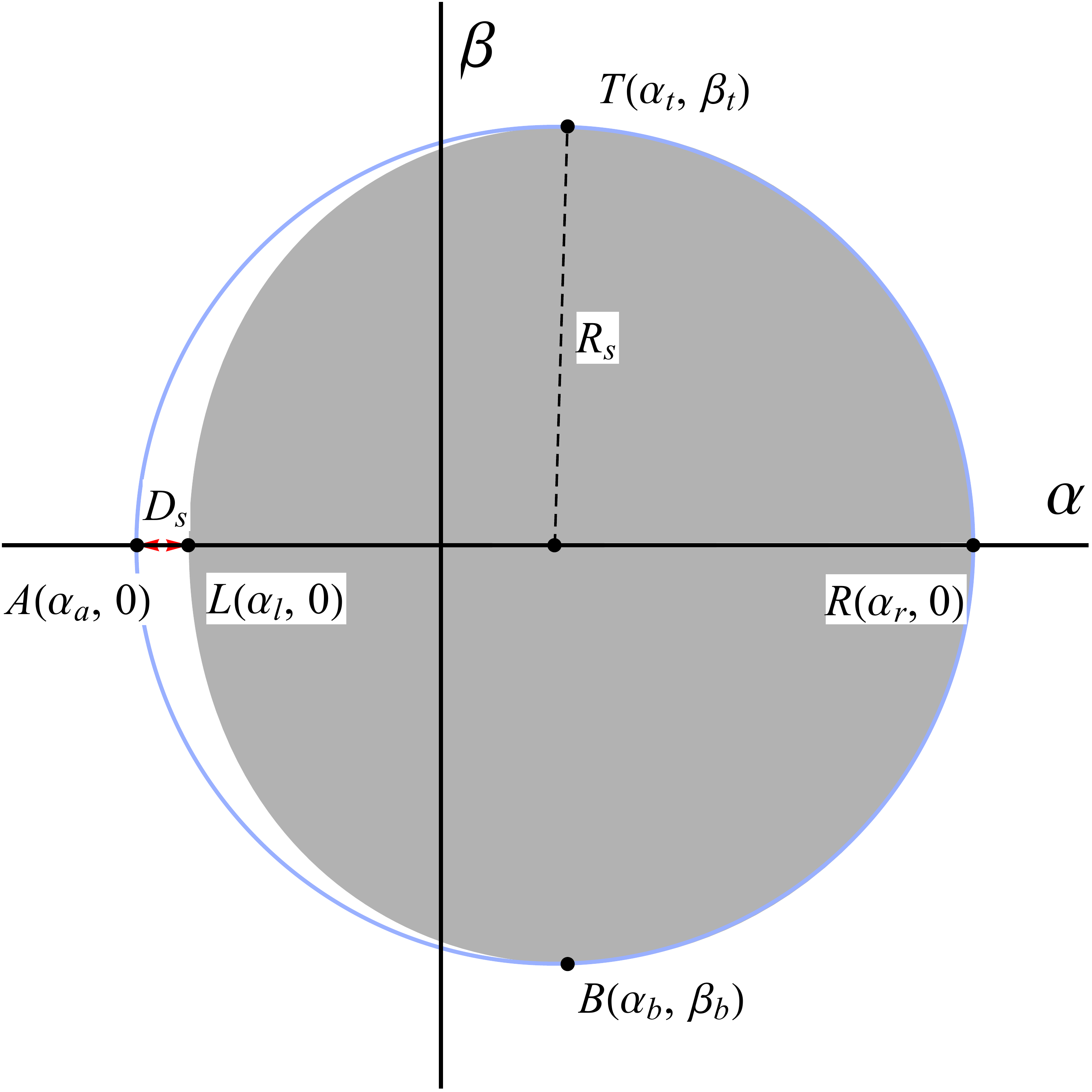}
  	\caption{Schematic illustration of the black hole shadow and the reference circle.}
  	\label{lRspic}
  \end{figure}

  Different shapes of the shadow can be obtained by plotting $\beta$ against $\alpha$. Fig. \ref{shadowofk} shows the plots of the shadows of the rotating nonlinear magnetic-charged black hole surrounded by PFDM for different values of the parameters $a$ and $k$. we found the size of the shadow increases with the increasing $\vert{k}\vert$ and the shadow gets more and more distorted with the increasing $a$. Interestingly, as shown in the last picture of Fig. \ref{shadowofk}, we found that when $\vert{k}\vert$ increases, the most left position (marked by $L$ in Fig. \ref{lRspic}) of the shadow will first move to the left and then to the right. As shown in Fig. \ref{most left}, we studied the dependence of the horizontal coordinate $\alpha_l$ of the most left position of the Shadow on the spin parameter $a$ and the PFDM intensity parameter $k$ with $Q=0.5$. In the case of $a=0.8$, we found that when $k\simeq{-1.23}$, the value of $\alpha_l$ is the smallest.

\begin{figure}[]
  	\centering
  	\includegraphics[width=.49\textwidth]{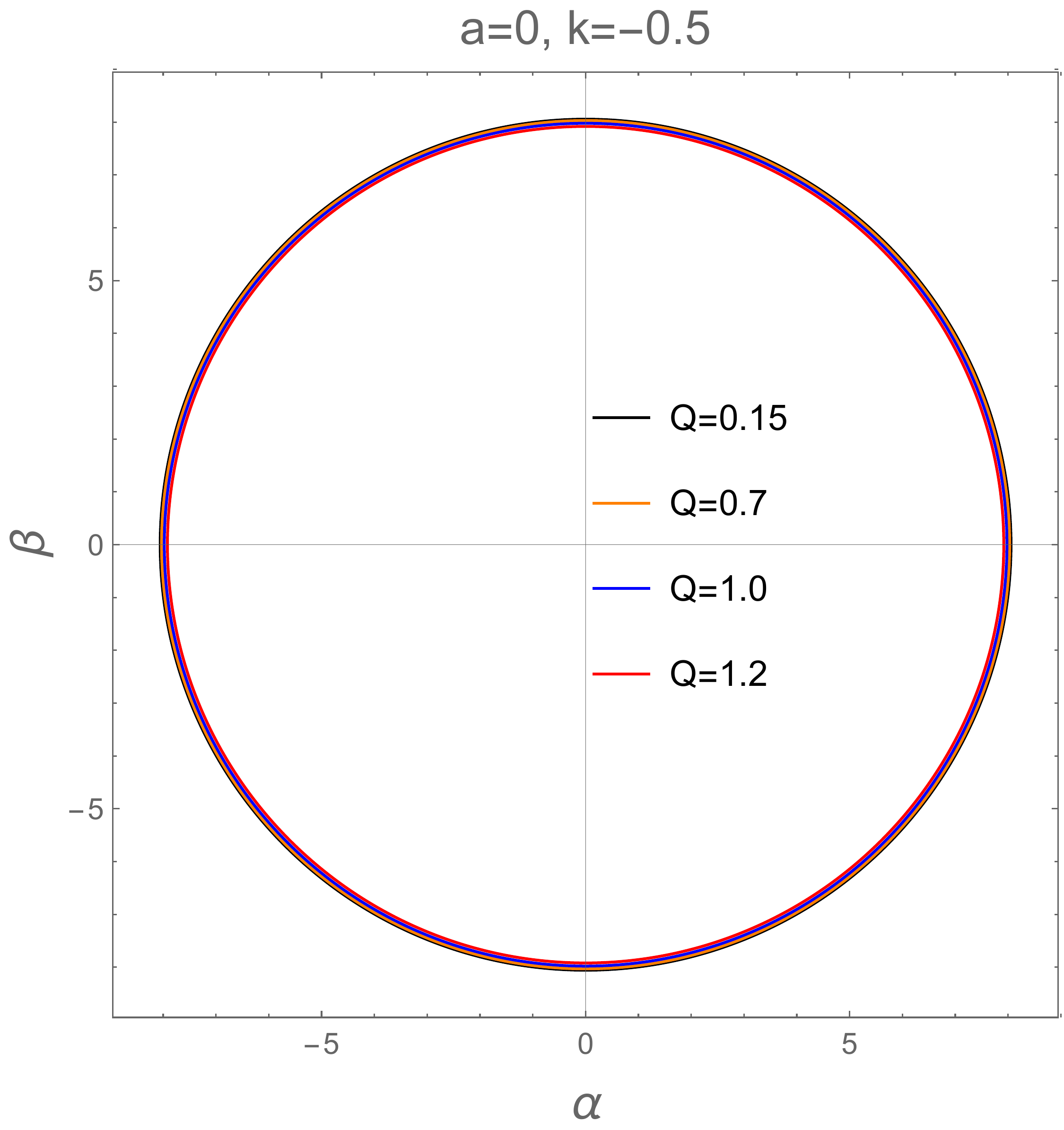}
  	\includegraphics[width=.49\textwidth]{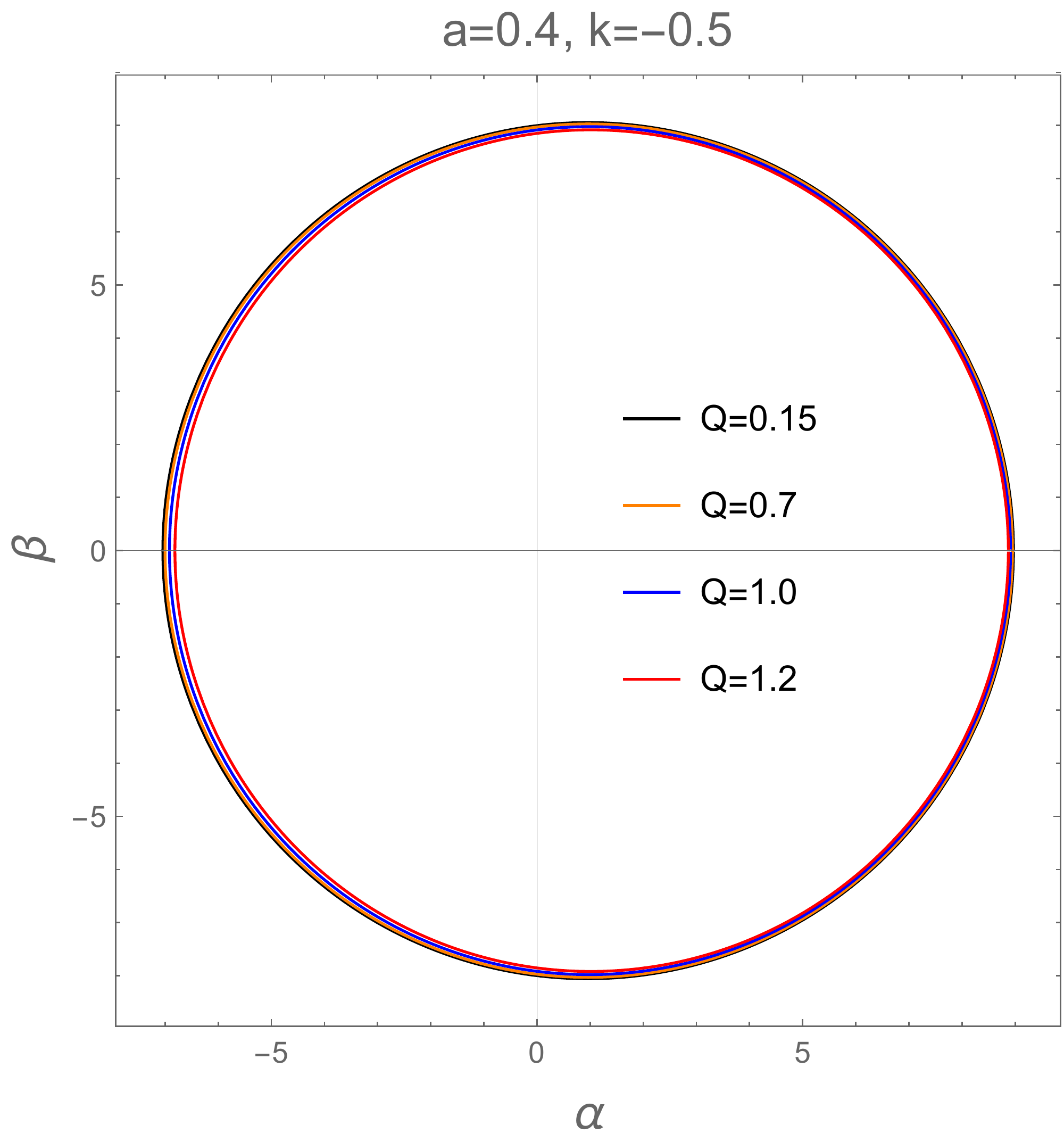}
  	\includegraphics[width=.49\textwidth]{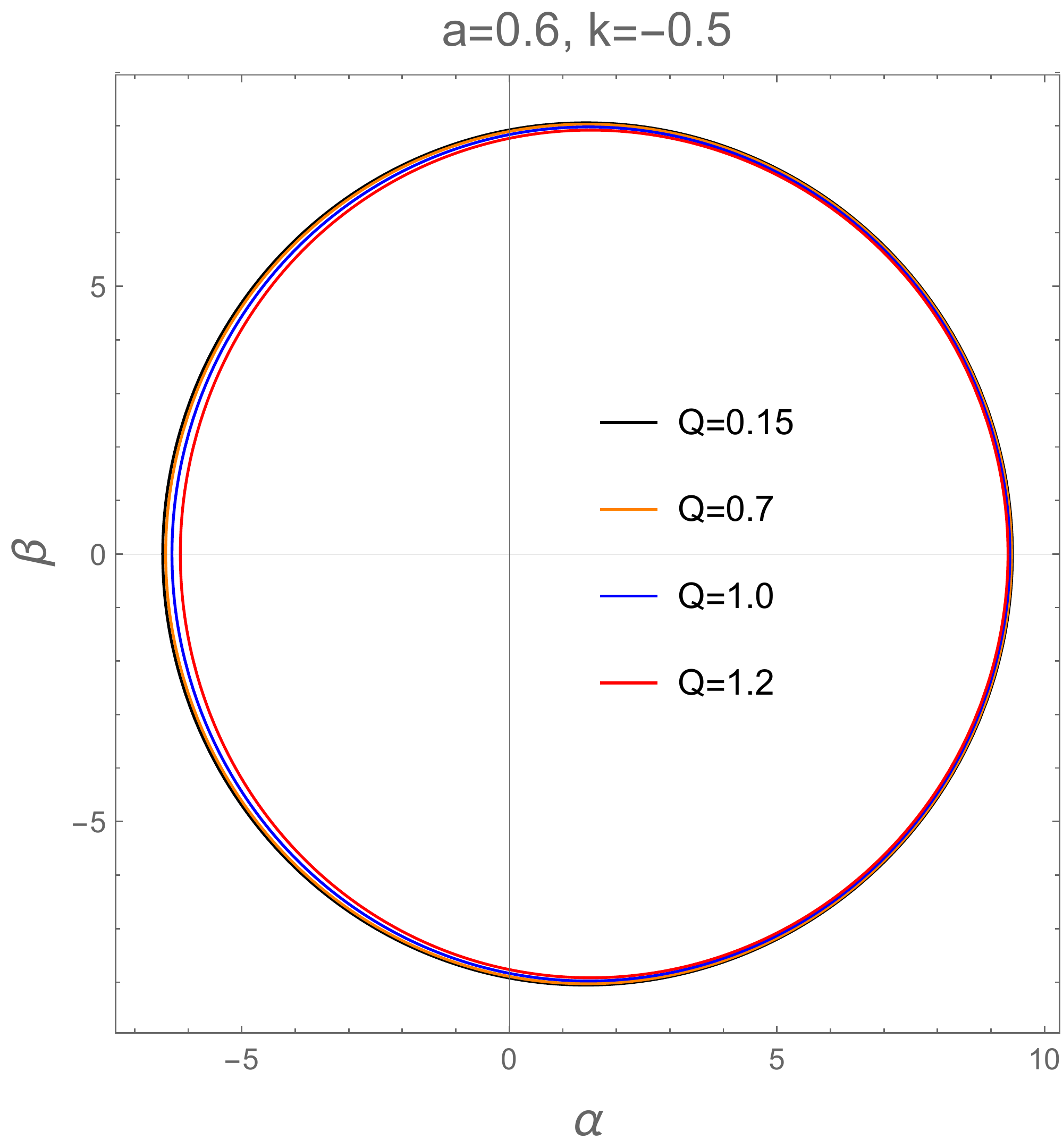}
  	\includegraphics[width=.49\textwidth]{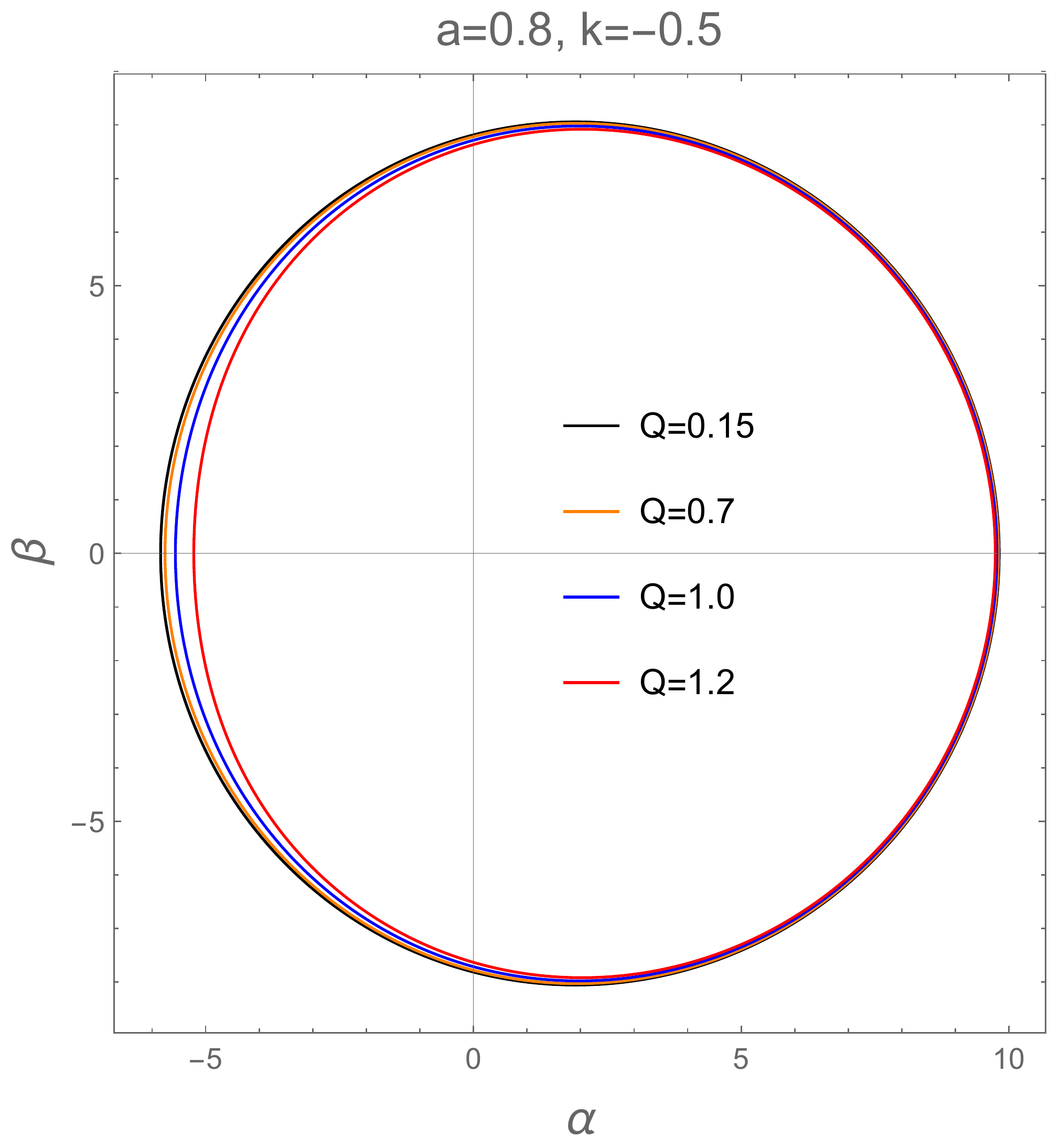}
  	\caption{Examples of shadows of the rotating nonlinear magnetic-charged black hole surrounded by PFDM for different values of the parameters $a$ and $Q$ with the PFDM intensity parameter $k=-0.5$  .}
  	\label{shadowofq}
  \end{figure}

  Fig. \ref{shadowofq} shows the plots of the shadows of the rotating nonlinear magnetic-charged black hole surrounded by PFDM for different values of the parameters $a$ and $Q$. We found the size of the shadow decreases with the increasing $Q$, but it is not obvious.

  \par
  In order to extract detailed information from the shadow and connect to astronomical observations, it is recommended to introduce two observables the radius $R_{s}$ and the distortion parameter $\delta_{s}$ \cite{hioki2009measurement}. As shown in Fig. \ref{lRspic}, $R_{s}$ is the radius of the reference circle passing through three points: the top one $T\left(\alpha_t,\beta_t \right) $, the bottom one $B\left(\alpha_b,\beta_b \right)$ and the most right one $R\left(\alpha_r,0 \right)$. The distortion parameter $\delta_s$ is defined as
  \begin{equation}\label{eq:deltads}
  \delta_{s}=\frac{D_s}{R_s}=\frac{\vert\alpha_l-{\alpha}_a\vert}{R_s},
  \end{equation}
  where $D_s$ is the distance from the most left position $\left(L \right) $ of the shadow to the reference circle $\left(A \right) $.
  The observable $R_s$ is defined as
  \begin{equation}\label{eq:Rds}
  R_{s}=\frac{\left( {\alpha}_t-{\alpha}_r\right)^{2}+{\beta}_{t}^{2} }{2{\vert{\alpha_r-\alpha_t}\vert}}.
  \end{equation}

\begin{figure}[]
  	\centering
  	\includegraphics[width=.49\textwidth]{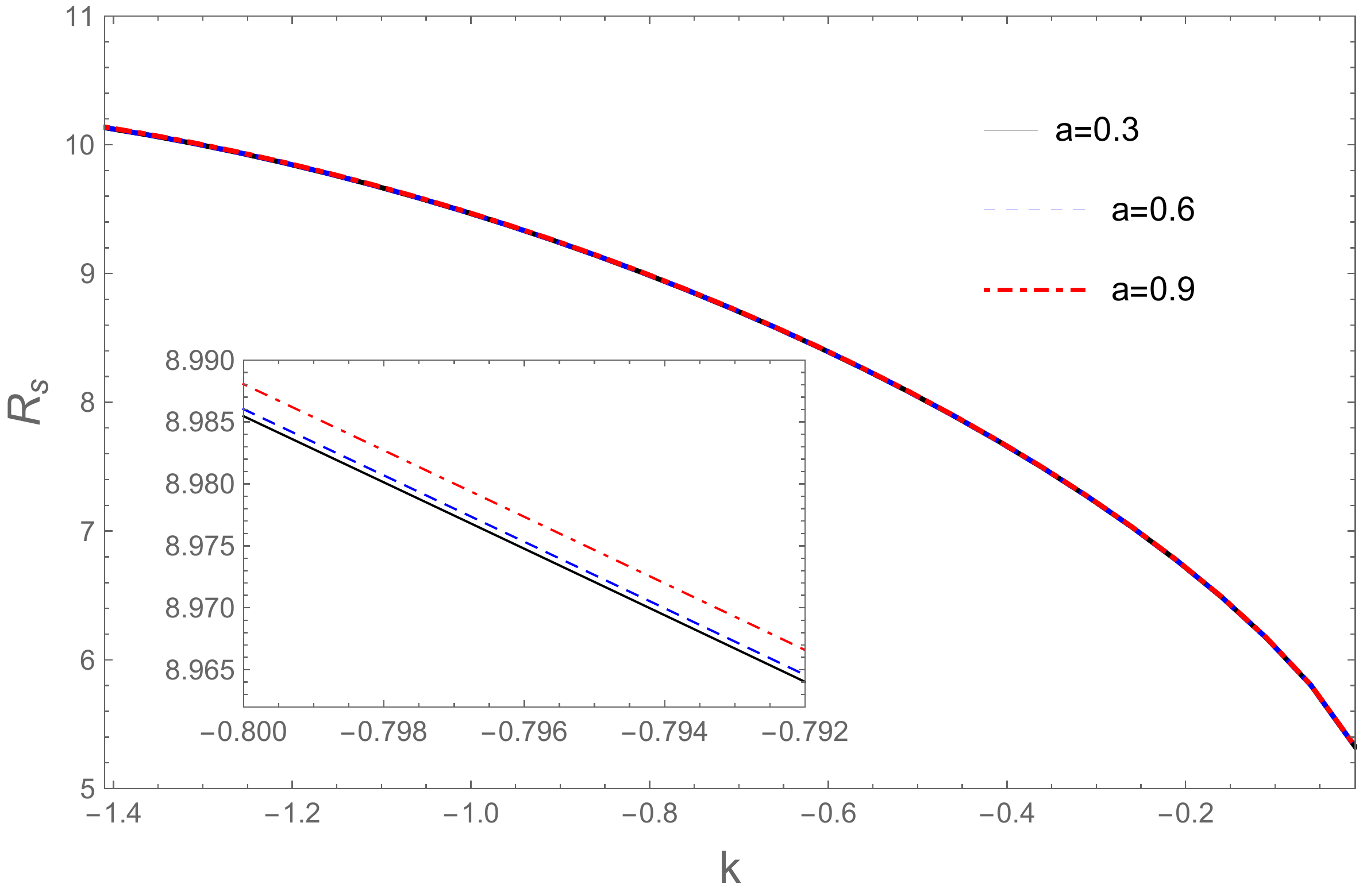}
  	\includegraphics[width=.49\textwidth]{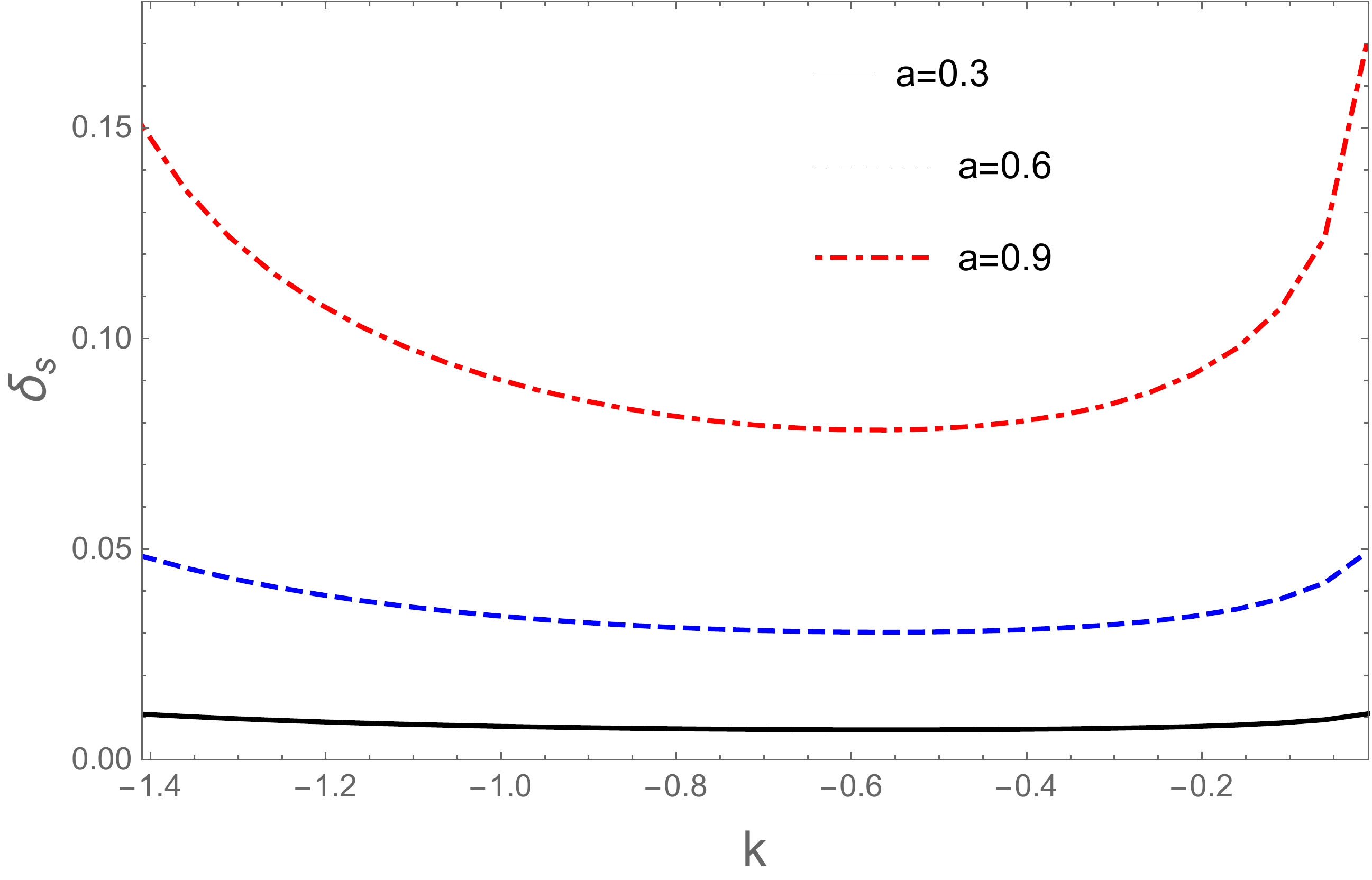}
  	\caption{The radius $R_{s}$ (left) and the distortion parameter $\delta_{s}$ (right) of the black hole shadow against the parameter $k$ for different values of the spin parameter $a$ with the magnetic charge $Q=0.5$ .}
  	\label{Rsk}
  \end{figure}

\begin{figure}[]
	\centering
	\includegraphics[width=.49\textwidth]{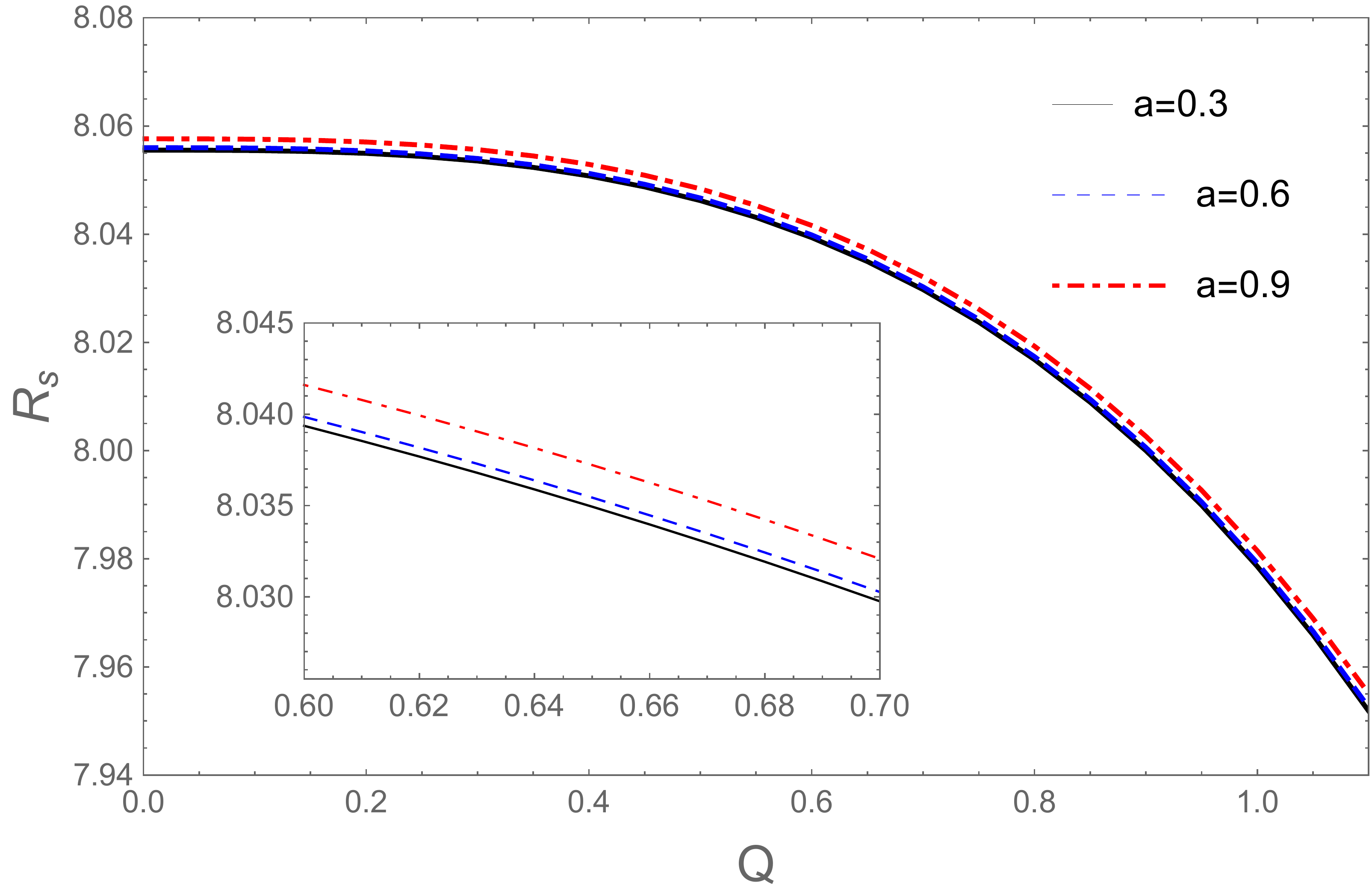}
	\includegraphics[width=.49\textwidth]{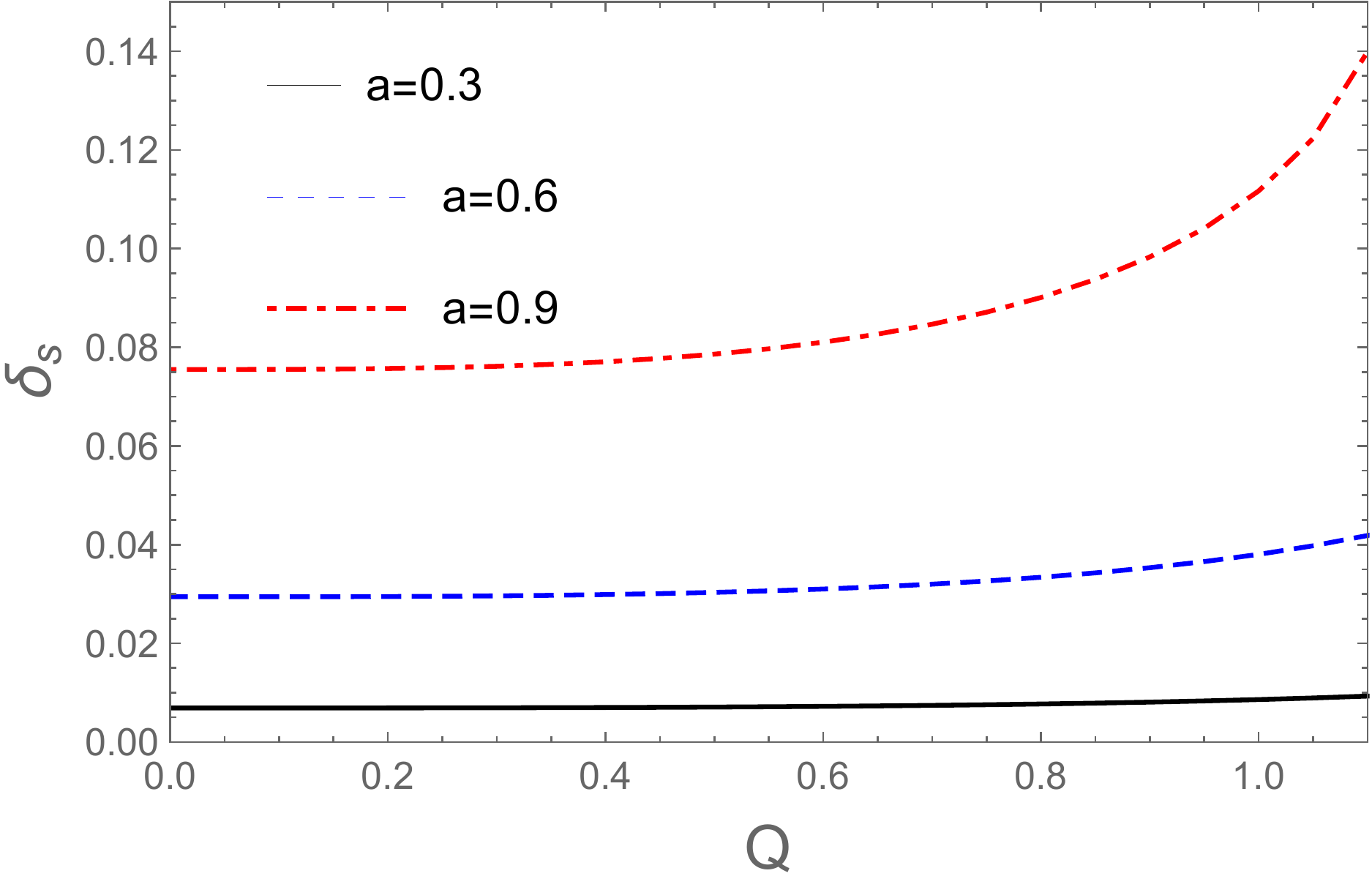}
	\caption{The radius $R_{s}$ (left) and the distortion parameter $\delta_{s}$ (right) of the black hole shadow against the parameter $Q$ for different values of the spin parameter $a$ with the PFDM intensity parameter $k=-0.5$ .}
	\label{Rsq}
  \end{figure}

  We show in Fig. \ref{Rsk} the evolution of the radius $R_s$ and the distortion parameter $\delta_s$ with the parameters $a$ and $k$.  We found the radius $R_s$ increases with the increasing $\vert{k}\vert$, this result is consistent with the result in Fig. \ref{shadowofk}. $\delta_s$ is not monotonic with respect to $k$. There exists a $k_{0}$ ($\simeq-0.56$). When $k<k_0$, $\delta_s$ increases with the increasing $\vert{k}\vert$. When $k>k_0$, $\delta_s$ decreases with the increasing $\vert{k}\vert$. Fig. \ref{Rsq} describes the behaviour of observables $R_s$ and $\delta_s$ with respect to the parameters $a$ and $Q$. We found that magnetic charge $Q$ diminishing the shadow radius $R_s$, while increases the distortion parameter $\delta_s$. From Figs. \ref{Rsk} and \ref{Rsq}, we also find that both $R_s$ and $\delta_s$ increase with the increasing spin parameter $a$.

  \section{Energy emission rate}\label{6}

  The high energy absorption cross section of a black hole oscillates around a limiting constant value $\sigma_{lim}$. For an observer at an infinite distance, the black hole shadow corresponds to the high energy absorption cross section.  Here, we compute the energy emission rate \cite{wei2013observing} using
  \begin{equation}
  \frac{\mathrm{d}^{2}E\left(\omega \right) }{\mathrm{d}\omega\mathrm{d}t}=\frac{2\pi^{2}\sigma_{lim}}{{e^{\omega/T}}-1}\omega^{3},
  \end{equation}
  the limiting constant value can be expressed as
  \begin{equation}
  \sigma_{lim}\approx\pi{R_s^{2}},
  \end{equation}
  $\omega$ represent the frequency of photon and $T$ is the Hawking temperature for the outer event horizon ($r_{+}$) which is defined by

  \begin{equation}
  T=\lim_{\theta=0,r\to{r_+}}\frac{\partial_{r}\sqrt{-g_{tt}}}{2\pi\sqrt{g_{rr}}}.
  \end{equation}
  In our case, for the rotating Hayward black hole in PFDM

  \begin{equation}
  \begin{split}
  g_{tt}=-\left(1-\frac{r^{2}-f(r)r^{2}}{\Sigma} \right),\quad
  g_{rr}=\frac{\Sigma}{\Delta_{r}}.
  \end{split}
  \end{equation}
  Thus, we obtain the Hawking temperature as
  \begin{equation}
  T=\frac{{r_+^2}f^{\prime}\left(r_{+} \right) \left(r_{+}^{2}+a^{2} \right)+2a^{2}r_{+}\left(f\left(r_{+} \right)-1  \right)  }{4\pi\left(r_{+}^{2}+a^{2} \right)^{2} }.
  \end{equation}

\begin{figure}[h]
  	\centering
  	\includegraphics[width=.32\textwidth]{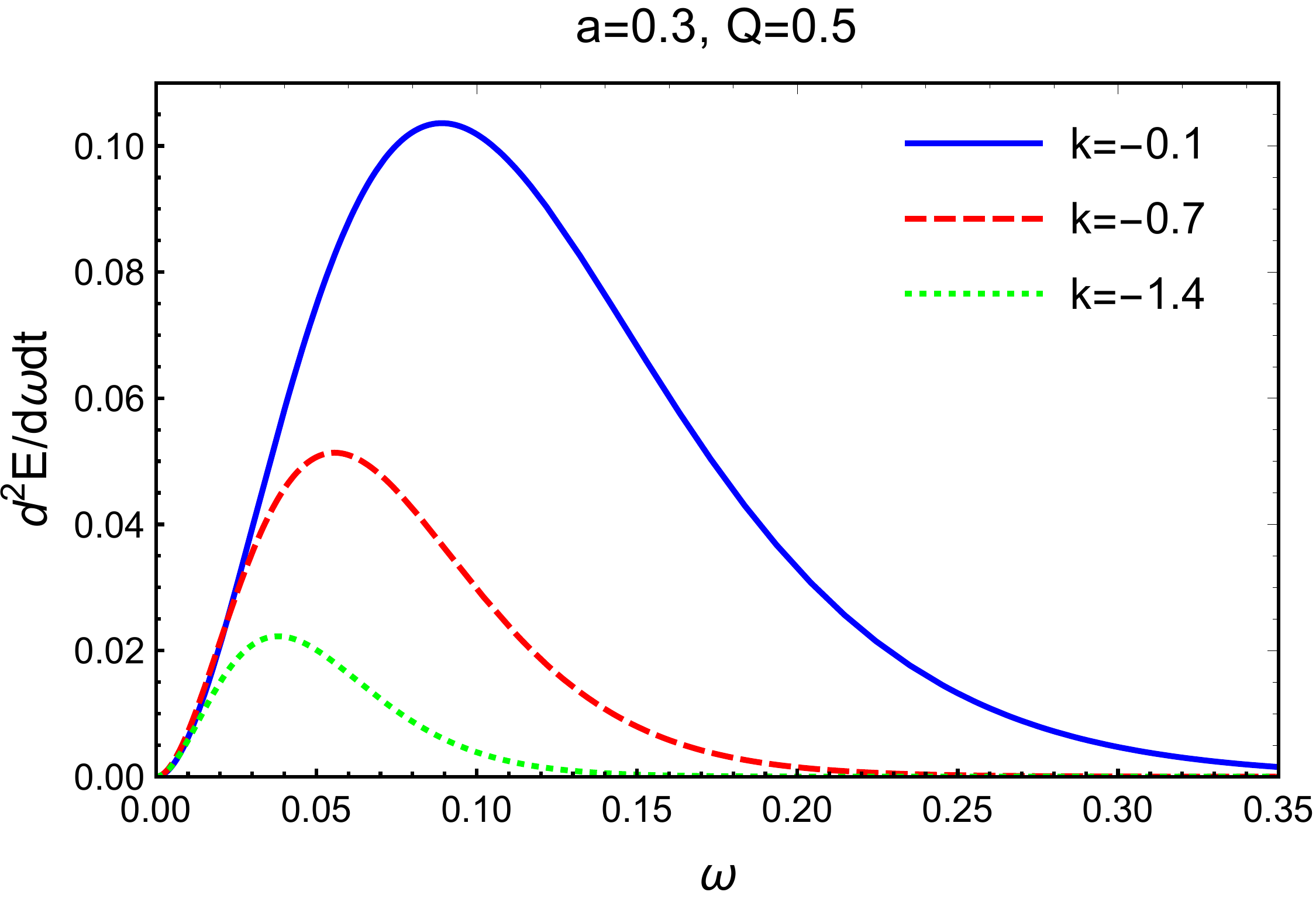}
  	\includegraphics[width=.32\textwidth]{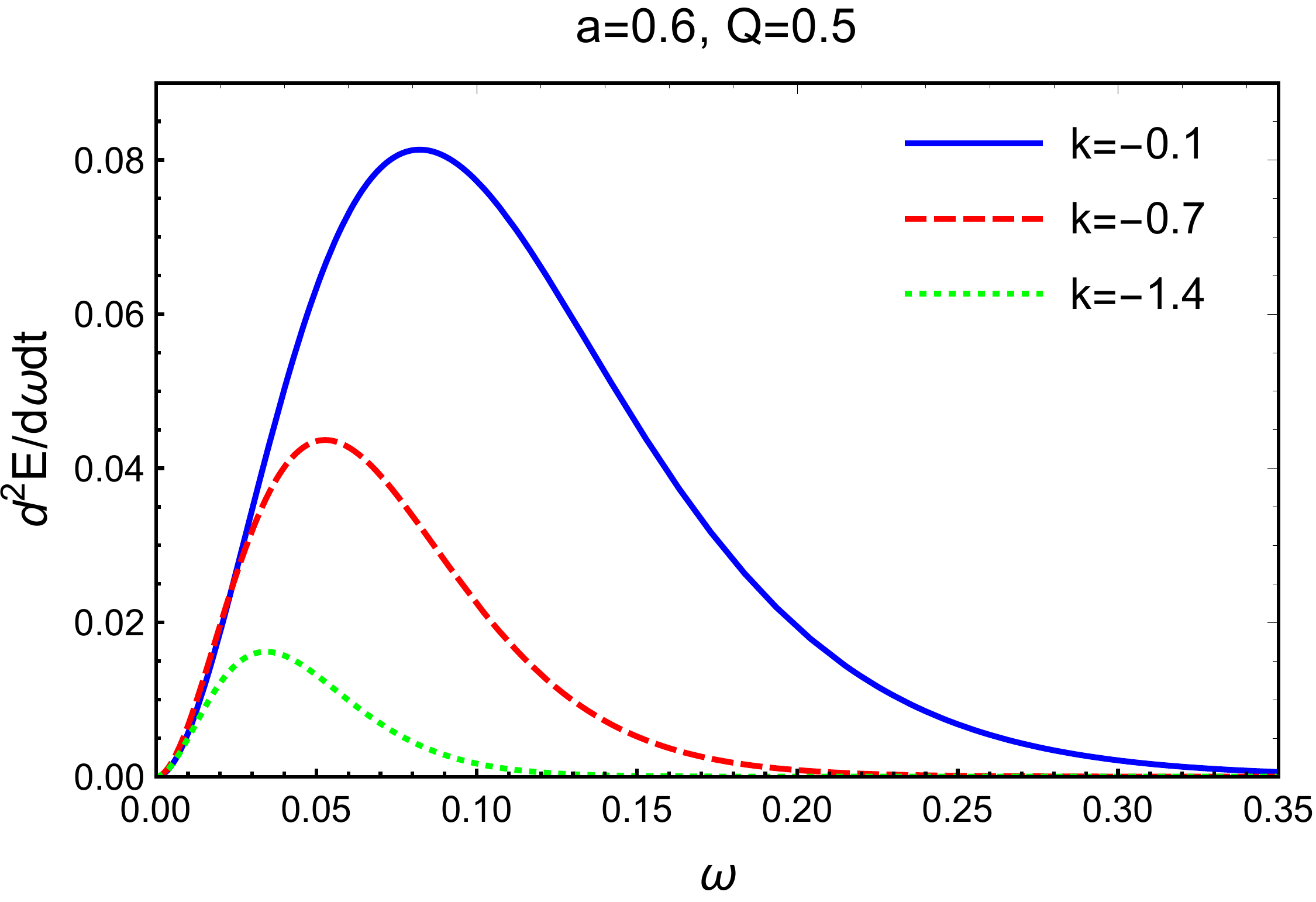}
  	\includegraphics[width=.32\textwidth]{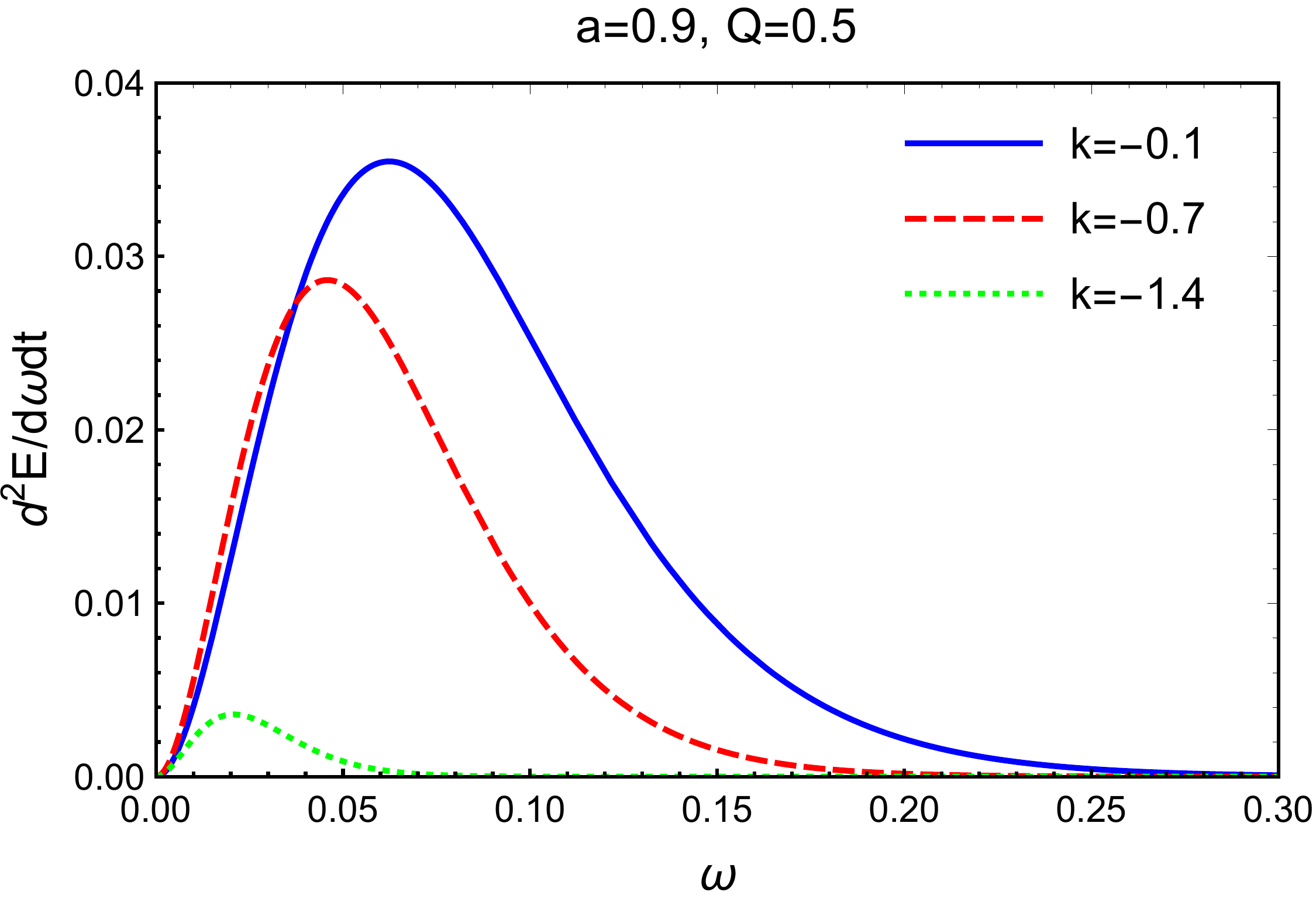}
  	\caption{ Evolution of the emission rate with the frequency $\omega$ for different values of the parameters $a$ and $k$ with the magnetic charge $Q=0.5$ .  }
  	\label{emissionk}
  \end{figure}

\begin{figure}[h]
	\centering
	\includegraphics[width=.32\textwidth]{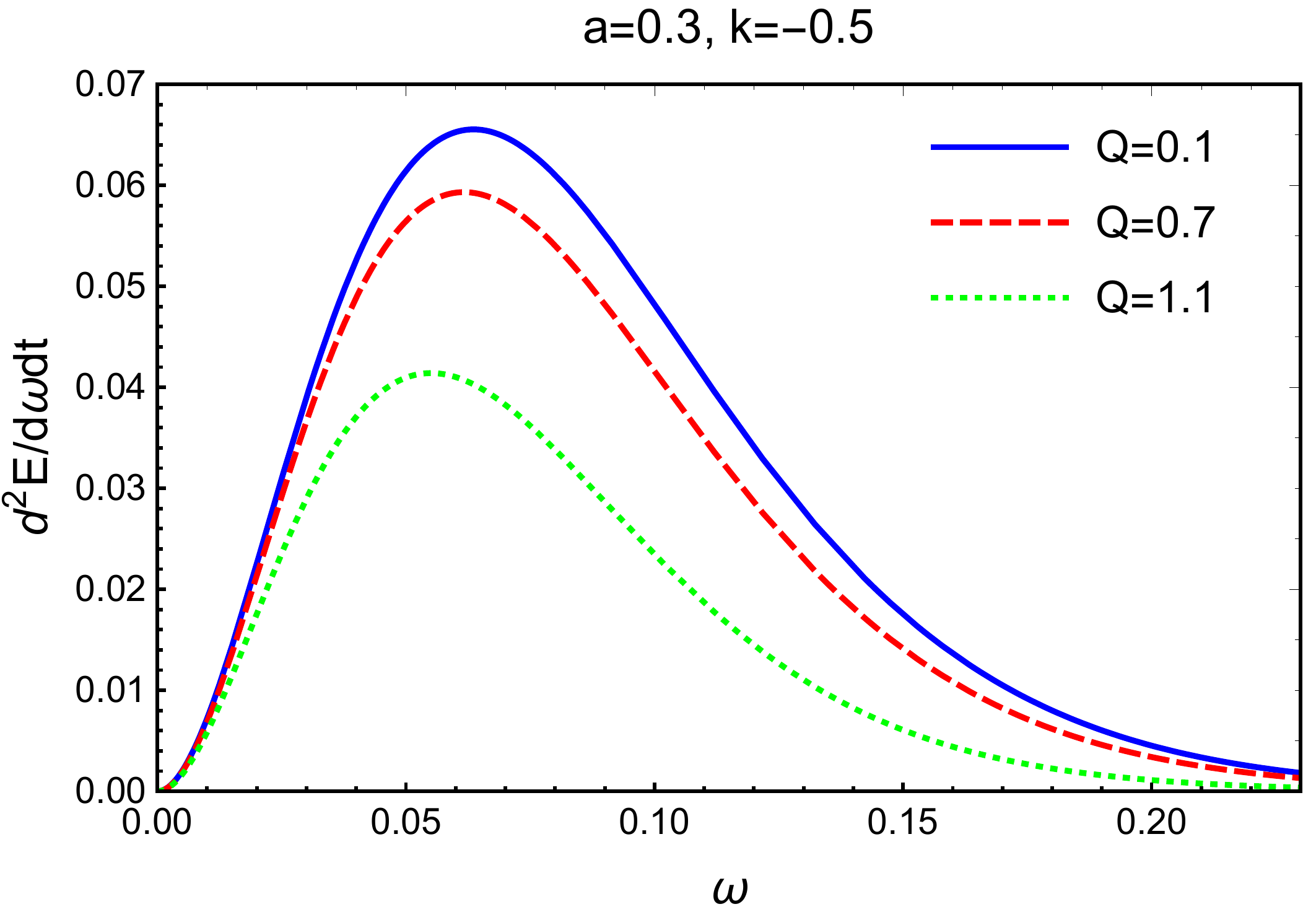}
	\includegraphics[width=.32\textwidth]{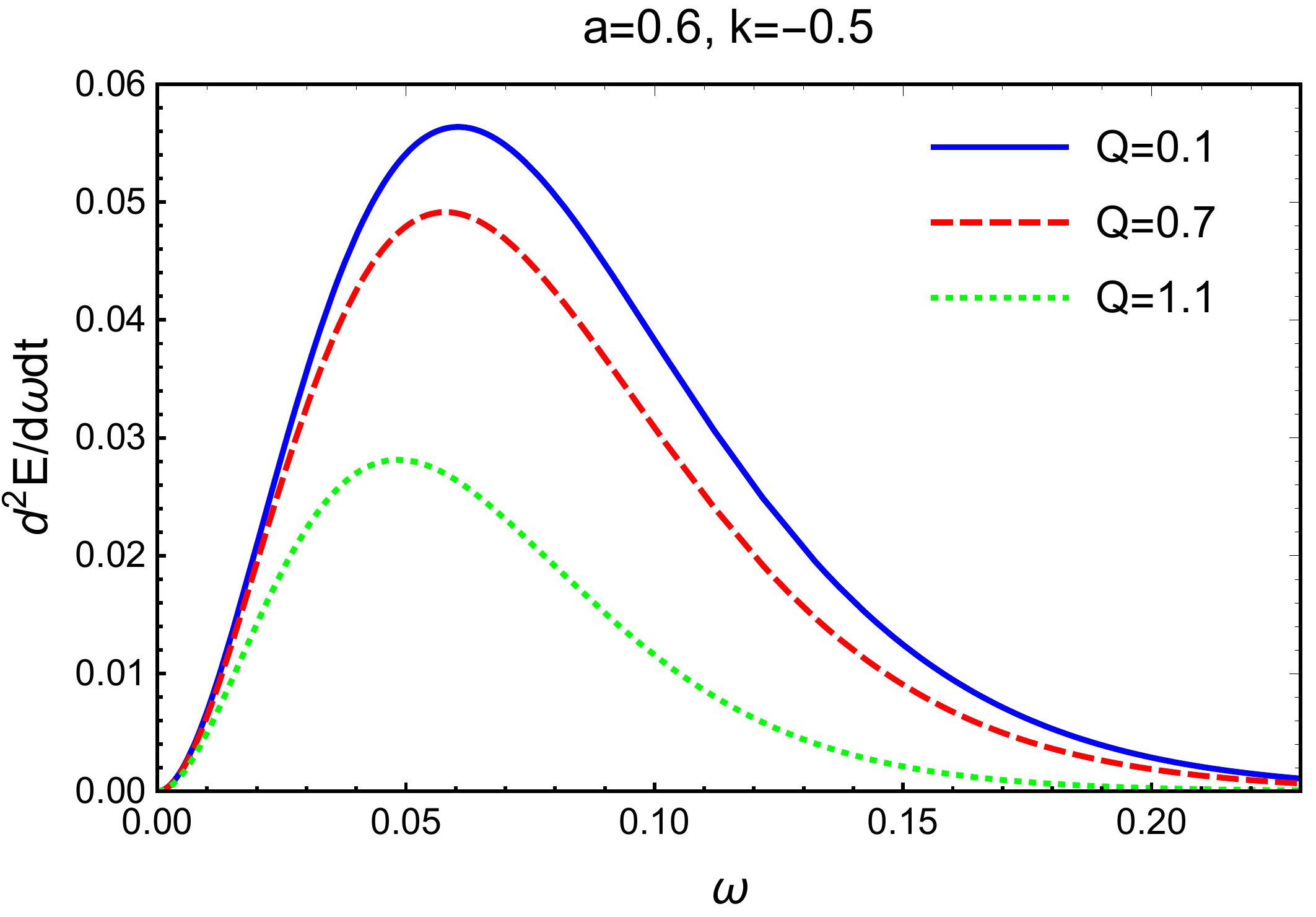}
	\includegraphics[width=.32\textwidth]{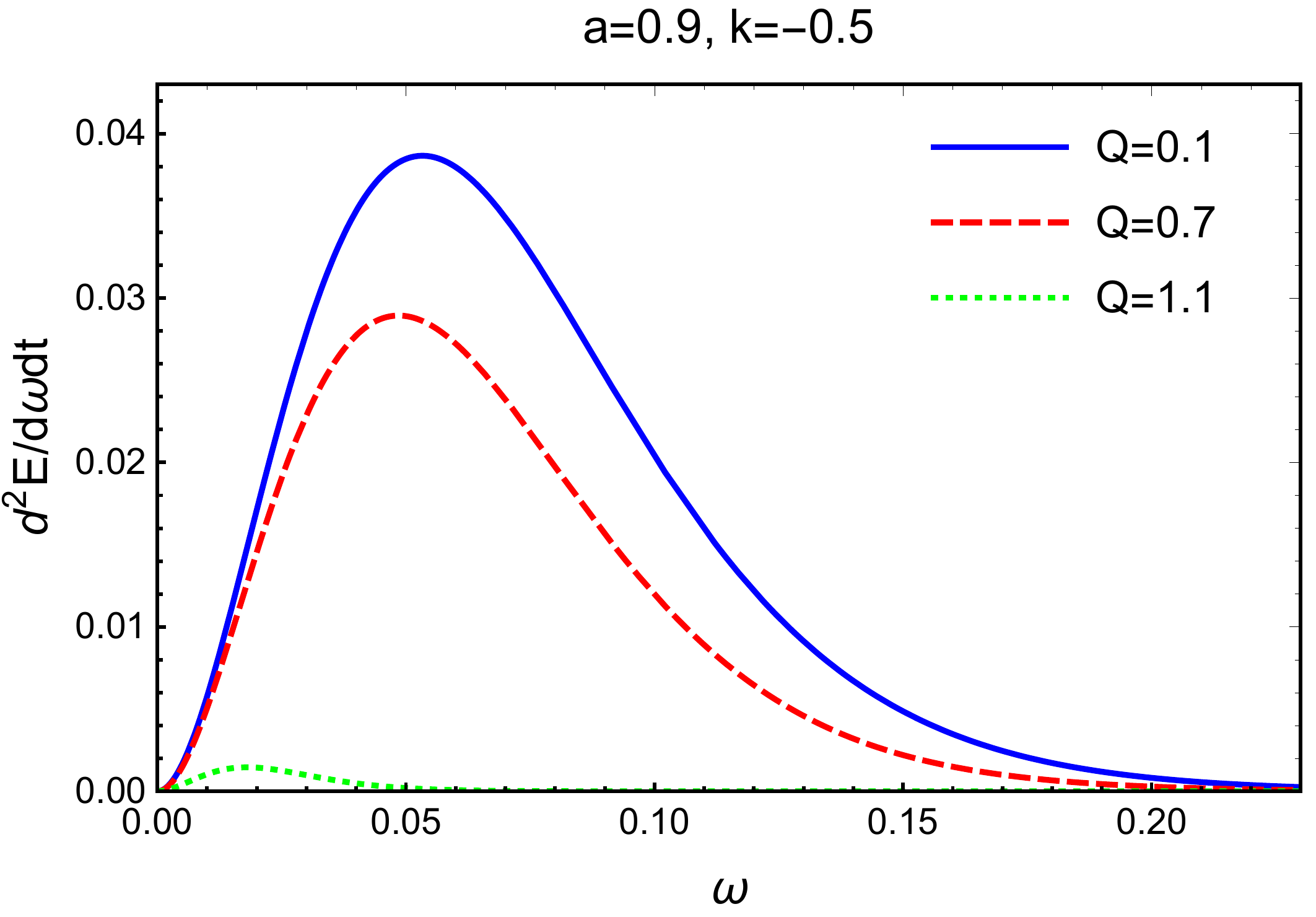}
	\caption{ Evolution of the emission rate with the frequency $\omega$ for different values of the parameters $a$ and $Q$ with the PFDM intensity parameter $k=-0.5$ .}
	\label{emissionq}
  \end{figure}

   We show energy emission rate in Fig. \ref{emissionk}, against the frequency $\omega$ for different values of the parameter $a$ and $k$. We can see that the peak of the emission decreases with increasing $\vert{k}\vert$ and shifts to left, i.e., to lower frequency. In Fig. \ref{emissionq}, we describe the energy emission rate against the frequency $\omega$ for different values of the parameter $a$ and $Q$. It is clear that the peak of the emission decreases with the increasing $Q$ and also shifts to lower frequency.

\section{Conclusions and discussions}\label{7}
  In this paper, we have obtained the exact solution of the static spherically symmetric Hayward black hole surrounded by perfect fluid dark matter and generalized it to the corresponding rotating metric with the Newman-Janis algorithm. We show that the photon regions of rotating Hayward black hole in PFDM can be divided into the stable regions and the unstable regions. Then we studied the black hole shadow. The shadow is a perfect circle in the non-rotating case ($a=0$) and a deformed one in the rotating case ($a\neq{0}$). We adopted two observables, the radius $R_s$ and the distortion parameter $\delta_s$, characterizing the apparent shape of the black hole shadow. For a fixed value of $a$, $R_s$ increases with the increasing $\vert{k}\vert$, but decreases with the increasing $Q$. There exists a $k_{0}$. When $k<k_0$, $\delta_s$ increases with the increasing $\vert{k}\vert$. When $k>k_0$, $\delta_s$ decreases with the increasing $\vert{k}\vert$. The black hole emission rate is also investigated, we find that the emission rate decreases with the increasing $\vert{k}\vert$ (or $Q$) and the peak of the emission shifts to lower frequency.

  \setlength{\tabcolsep}{3pt}
  \begin{table*}
  	\centering
  	\begin{tabular}{lccccccccccccccc}
  		\hline
  		${}$ & &   \multicolumn{3}{c}{} & & &\multicolumn{3}{c}{a=0.1}  & & & \multicolumn{3}{c}{}  \\
  		\hline
  		$Q$ & &   \multicolumn{3}{c}{0.3} & & &\multicolumn{3}{c}{0.6}  & & & \multicolumn{3}{c}{0.9 }  \\
  		\hline
  		$k$ & &$-0.3$  &$-0.6$  &$-0.9$  & & &$-0.3$  &$-0.6$  &$-0.9$   & & &$-0.3$  &$-0.6$  &$-0.9$\\
  		\hline
  		$R_s$ & &  7.2316 & 8.3992  &9.2450  & & &7.2144  &8.3859  &9.2329   & & &7.1662  &8.3490  &9.1994  \\
  		$\delta_s $ (\%) & &0.0790 & 0.0758 &0.0812  & & &0.0832 &0.0785 &0.0836       & & &0.0964 &0.0868 & 0.0908 \\
  		$\theta_s$ & &36.981  &42.953  &47.278   & & &36.894  &42.885   &47.216   & & &36.647  &42.696  &47.045  \\
  		\hline
  		${}$ & &   \multicolumn{3}{c}{} & & &\multicolumn{3}{c}{a=0.8}  & & & \multicolumn{3}{c}{}  \\
  		\hline
  		$Q$ & &   \multicolumn{3}{c}{0.3} & & &\multicolumn{3}{c}{0.6}  & & & \multicolumn{3}{c}{0.9 }  \\
  		\hline
  		$k$ & &$-0.3$  &$-0.6$  &$-0.9$  & & &$-0.3$  &$-0.6$  &$-0.9$   & & &$-0.3$  &$-0.6$  &$-0.9$\\
  		\hline
  		$R_s$ & &  7.2329 & 8.4007  &9.2468  & & &7.2158  &8.3875  &9.2348   & & &7.1680  &8.3509  &9.2016  \\
  		$\delta_s $ (\%) & &6.0122 & 5.6908 &6.1653  & & &6.4817 &5.9762 &6.4335       & & &8.1835 &6.9034 & 7.2959 \\
  		$\theta_s$ & &36.988  &42.960  &47.287   & & &36.901  &42.893   &47.226   & & &36.656  &42.705  &47.056  \\
  		\hline
  	\end{tabular}
  	\caption{The radius $R_{s}$, the distortion parameter $\delta_{s}$ and the angular size $\theta_s$ for the supermassive black hole $\mathrm{Sgr\ A^{*}}$ at the center of the Milky Way, for different values of black hole spin $a$, magnetic charge $Q$ and PFDM parameter $k$.}
  	\flushleft
  	\label{thetas}
  \end{table*}

  We can also use the shadow radius $R_s$ to determine the angular size of the black hole. The angular size is given by $\theta_{s}=R_{s}M/D_{o}$, with $M$ the black hole mass and $D_{o}$ the distance from the observer to the black hole \cite{amarilla2012shadow}. The angular size can be further expressed as $\theta_s=9.87098\times{10}^{-6}R_{s}\left(M/M_{\odot} \right)\left(1 \mathrm{kpc}/D_{o} \right)\ \mu{\mathrm{as}}  $ with $M_{\odot}$ the solar mass. For the supermassive black hole $\mathrm{Sgr\ A^{*}}$ at the center of the Milky Way, $M=4.3\times{10}^{6}M_{\odot}$ and $D_{o}=8.3\ \mathrm{kpc}$, we show the calculation result in table \ref{thetas}. From the table, we find that to extract the information of the black hole spin $a$ or magnetic charge $Q$, the required angular resolution must be less than $1\ \mu{\mathrm{as}}$. However, for PFDM parameter $k$, a resolution of $1\ \mu{\mathrm{as}}$ is sufficient. Currently, EHT has a resolution of $\ \sim{60}\ \mu\mathrm{as}$ at 230 GHz. In the future, the spacebased VLBI RadioAstron \cite{kardashev2013radioastron} will obtain a resolution of $\ \sim{1-10}\ \mu\mathrm{as}$. We expect that future observations can distinguish the influence of PFDM on the shadow of the black hole $\mathrm{Sgr\ A^{*}}$.

\section*{Conflicts of Interest}
  The authors declare that there are no conflicts of interest regarding the publication of this paper.

\section*{Acknowledgments}
  We would like to thank the National Natural Science Foundation of China (Grant No.11571342) for supporting us on this work.
  This work makes use of the Black Hole Perturbation Toolkit.

\section*{References}

 \bibliographystyle{unsrt}
 \bibliography{ref}

\begin{thebibliography}{10}

\bibitem{abbott2016gw150914}
Benjamin~P Abbott, R~Abbott, TD~Abbott, MR~Abernathy, F~Acernese, K~Ackley,
  C~Adams, T~Adams, P~Addesso, RX~Adhikari, et~al.
\newblock {GW150914: The Advanced LIGO detectors in the era of first
  discoveries}.
\newblock {\em Physical review letters}, 116(13):131103, 2016.

\bibitem{event2019first}
Event Horizon~Telescope Collaboration et~al.
\newblock First m87 event horizon telescope results. i. the shadow of the
  supermassive black hole.
\newblock {\em arXiv preprint arXiv:1906.11238}, 2019.

\bibitem{akiyama2019first}
Kazunori Akiyama, Antxon Alberdi, Walter Alef, Keiichi Asada, Rebecca Azulay,
  Anne-Kathrin Baczko, David Ball, Mislav Balokovi{\'c}, John Barrett, Dan
  Bintley, et~al.
\newblock First m87 event horizon telescope results. iv. imaging the central
  supermassive black hole.
\newblock {\em The Astrophysical Journal Letters}, 875(1):L4, 2019.

\bibitem{synge1966escape}
JL~Synge.
\newblock The escape of photons from gravitationally intense stars.
\newblock {\em Monthly Notices of the Royal Astronomical Society},
  131(3):463--466, 1966.

\bibitem{bardeen1973inblack}
JM~Bardeen.
\newblock inblack holes, edited by c. dewitt and b. dewitt, 1973.

\bibitem{grenzebach2014photon}
Arne Grenzebach, Volker Perlick, and Claus L{\"a}mmerzahl.
\newblock Photon regions and shadows of kerr-newman-nut black holes with a
  cosmological constant.
\newblock {\em Physical Review D}, 89(12):124004, 2014.

\bibitem{stuchlik2018light}
Zden{\v{e}}k Stuchl{\'\i}k, Daniel Charbul{\'a}k, and Jan Schee.
\newblock Light escape cones in local reference frames of kerr--de sitter black
  hole spacetimes and related black hole shadows.
\newblock {\em The European Physical Journal C}, 78(3):1--32, 2018.

\bibitem{li2020shadow}
Peng-Cheng Li, Minyong Guo, and Bin Chen.
\newblock Shadow of a spinning black hole in an expanding universe.
\newblock {\em Physical Review D}, 101(8):084041, 2020.

\bibitem{atamurotov2013shadow}
Farruh Atamurotov, Ahmadjon Abdujabbarov, and Bobomurat Ahmedov.
\newblock Shadow of rotating non-kerr black hole.
\newblock {\em Physical Review D}, 88(6):064004, 2013.

\bibitem{grenzebach2015photon}
Arne Grenzebach, Volker Perlick, and Claus L{\"a}mmerzahl.
\newblock Photon regions and shadows of accelerated black holes.
\newblock {\em International Journal of Modern Physics D}, 24(09):1542024,
  2015.

\bibitem{vagnozzi2019hunting}
Sunny Vagnozzi and Luca Visinelli.
\newblock Hunting for extra dimensions in the shadow of m87.
\newblock {\em Physical Review D}, 100(2):024020, 2019.

\bibitem{banerjee2020silhouette}
Indrani Banerjee, Sumanta Chakraborty, and Soumitra SenGupta.
\newblock Silhouette of m87*: A new window to peek into the world of hidden
  dimensions.
\newblock {\em Physical Review D}, 101(4):041301, 2020.

\bibitem{ahmed20205d}
Fazlay Ahmed, Dharm~Veer Singh, and Sushant~G Ghosh.
\newblock 5d rotating regular myers-perry black holes and their shadow.
\newblock {\em arXiv preprint arXiv:2008.10241}, 2020.

\bibitem{chang2020black}
Zhe Chang and Qing-Hua Zhu.
\newblock Black hole shadow in the view of freely falling observers.
\newblock {\em Journal of Cosmology and Astroparticle Physics}, 2020(06):055,
  2020.

\bibitem{contreras2019black}
Ernesto Contreras, JM~Ramirez-Velasquez, {\'A}ngel Rinc{\'o}n, Grigoris
  Panotopoulos, and Pedro Bargue{\~n}o.
\newblock Black hole shadow of a rotating polytropic black hole by the
  newman--janis algorithm without complexification.
\newblock {\em The European Physical Journal C}, 79(9):802, 2019.

\bibitem{belhaj2020black}
A~Belhaj, M~Benali, A~El Balali, W~El Hadri, H~El Moumni, and E~Torrente-Lujan.
\newblock Black hole shadows in m-theory scenarios.
\newblock {\em arXiv preprint arXiv:2008.09908}, 2020.

\bibitem{gralla2019black}
Samuel~E Gralla, Daniel~E Holz, and Robert~M Wald.
\newblock Black hole shadows, photon rings, and lensing rings.
\newblock {\em Physical Review D}, 100(2):024018, 2019.

\bibitem{jusufi2020connection}
Kimet Jusufi.
\newblock Connection between the shadow radius and quasinormal modes in
  rotating spacetimes.
\newblock {\em arXiv preprint arXiv:2004.04664}, 2020.

\bibitem{jusufi2020constraining}
Kimet Jusufi, Mustapha Azreg-A{\"\i}nou, and Mubasher Jamil.
\newblock Constraining the generalized uncertainty principle: Black hole shadow
  m87* and quasiperiodic oscillations.
\newblock {\em arXiv preprint arXiv:2008.09115}, 2020.

\bibitem{zaman2020optical}
Gulmina Zaman~Babar, Abdullah Zaman~Babar, and Farruh Atamurotov.
\newblock Optical properties of kerr-newman spacetime in the presence of
  plasma.
\newblock {\em arXiv e-prints}, pages arXiv--2008, 2020.

\bibitem{uccanok2020orbits}
Onur U{\c{c}}anok.
\newblock Orbits around a kerr black hole and its shadow.
\newblock {\em arXiv preprint arXiv:2002.07057}, 2020.

\bibitem{eiroa2018shadow}
Ernesto~F Eiroa and Carlos~M Sendra.
\newblock Shadow cast by rotating braneworld black holes with a cosmological
  constant.
\newblock {\em The European Physical Journal C}, 78(2):1--8, 2018.

\bibitem{maceda2020shadow}
Marco Maceda, Alfredo Mac{\'\i}as, and Daniel Mart{\'\i}nez-Carbajal.
\newblock Shadow of a noncommutative inspired einstein-euler-heisenberg black
  hole.
\newblock {\em arXiv preprint arXiv:2008.07040}, 2020.

\bibitem{abdujabbarov2016shadow}
Ahmadjon Abdujabbarov, Muhammed Amir, Bobomurat Ahmedov, and Sushant~G Ghosh.
\newblock Shadow of rotating regular black holes.
\newblock {\em Physical Review D}, 93(10):104004, 2016.

\bibitem{chen2020optical}
Yuan Chen, He-Xu Zhang, Tian-Chi Ma, and Jian-Bo Deng.
\newblock Optical properties of a nonlinear magnetic charged rotating black
  hole surrounded by quintessence with a cosmological constant.
\newblock {\em arXiv preprint arXiv:2009.03778}, 2020.

\bibitem{he2020shadows}
Peng-Zhang He, Qi-Qi Fan, Hao-Ran Zhang, and Jian-Bo Deng.
\newblock Shadows of rotating hayward-de sitter black holes with astrometric
  observables.
\newblock {\em arXiv preprint arXiv:2009.06705}, 2020.

\bibitem{cunha2018shadows}
Pedro~VP Cunha and Carlos~AR Herdeiro.
\newblock Shadows and strong gravitational lensing: a brief review.
\newblock {\em General Relativity and Gravitation}, 50(4):42, 2018.

\bibitem{cunha2015shadows}
Pedro~VP Cunha, Carlos~AR Herdeiro, Eugen Radu, and Helgi~F R{\'u}narsson.
\newblock Shadows of kerr black holes with scalar hair.
\newblock {\em Physical review letters}, 115(21):211102, 2015.

\bibitem{cunha2020stationary}
Pedro~VP Cunha and Carlos~AR Herdeiro.
\newblock Stationary black holes and light rings.
\newblock {\em Physical Review Letters}, 124(18):181101, 2020.

\bibitem{dokuchaev2020visible}
Vyacheslav~I Dokuchaev and Natalia~O Nazarova.
\newblock Visible shapes of black holes m87* and sgra.
\newblock {\em arXiv preprint arXiv:2007.14121}, 2020.

\bibitem{allahyari2020magnetically}
Alireza Allahyari, Mohsen Khodadi, Sunny Vagnozzi, and David~F Mota.
\newblock Magnetically charged black holes from non-linear electrodynamics and
  the event horizon telescope.
\newblock {\em Journal of Cosmology and Astroparticle Physics}, 2020(02):003,
  2020.

\bibitem{khodadi2020black}
Mohsen Khodadi, Alireza Allahyari, Sunny Vagnozzi, and David~F Mota.
\newblock Black holes with scalar hair in light of the event horizon telescope.
\newblock {\em arXiv preprint arXiv:2005.05992}, 2020.

\bibitem{badia2020influence}
Javier Bad{\'\i}a and Ernesto~F Eiroa.
\newblock The influence of an anisotropic matter field on the shadow of a
  rotating black hole.
\newblock {\em arXiv preprint arXiv:2005.03690}, 2020.

\bibitem{zhang2020optical}
He-Xu Zhang, Cong Li, Peng-Zhang He, Qi-Qi Fan, and Jian-Bo Deng.
\newblock Optical properties of a brane-world black hole as photons couple to
  the weyl tensor.
\newblock {\em European Physical Journal C}, 80(5):1--11, 2020.

\bibitem{bardeen1968non}
James~M Bardeen.
\newblock Non-singular general-relativistic gravitational collapse.
\newblock In {\em Proc. Int. Conf. GR5, Tbilisi}, volume 174, 1968.

\bibitem{ayon1998regular}
Eloy Ayon-Beato and Alberto Garcia.
\newblock Regular black hole in general relativity coupled to nonlinear
  electrodynamics.
\newblock {\em Physical review letters}, 80(23):5056, 1998.

\bibitem{berej2006regular}
Waldemar Berej, Jerzy Matyjasek, Dariusz Tryniecki, and Mariusz Woronowicz.
\newblock Regular black holes in quadratic gravity.
\newblock {\em General Relativity and Gravitation}, 38(5):885--906, 2006.

\bibitem{hayward2006formation}
Sean~A Hayward.
\newblock Formation and evaporation of nonsingular black holes.
\newblock {\em Physical review letters}, 96(3):031103, 2006.

\bibitem{ade2016planck}
Peter~AR Ade, N~Aghanim, M~Arnaud, Mark Ashdown, J~Aumont, C~Baccigalupi,
  AJ~Banday, RB~Barreiro, JG~Bartlett, N~Bartolo, et~al.
\newblock Planck 2015 results-xiii. cosmological parameters.
\newblock {\em Astronomy \& Astrophysics}, 594:A13, 2016.

\bibitem{kiselev2003quintessence}
VV~Kiselev.
\newblock Quintessence and black holes.
\newblock {\em Classical and Quantum Gravity}, 20(6):1187, 2003.

\bibitem{toshmatov2017rotating}
Bobir Toshmatov, Zden{\v{e}}k Stuchl{\'\i}k, and Bobomurat Ahmedov.
\newblock Rotating black hole solutions with quintessential energy.
\newblock {\em The European Physical Journal Plus}, 132(2):1--21, 2017.

\bibitem{benavides2020rotating}
Carlos~A Benavides-Gallego, Ahmadjon Abdujabbarov, and Cosimo Bambi.
\newblock Rotating and nonlinear magnetic-charged black hole surrounded by
  quintessence.
\newblock {\em Physical Review D}, 101(4):044038, 2020.

\bibitem{ghosh2018lovelock}
Sushant~G Ghosh, Sunil~D Maharaj, Dharmanand Baboolal, and Tae-Hun Lee.
\newblock Lovelock black holes surrounded by quintessence.
\newblock {\em The European Physical Journal C}, 78(2):1--8, 2018.

\bibitem{ghosh2016rotating}
Sushant~G Ghosh.
\newblock Rotating black hole and quintessence.
\newblock {\em The European Physical Journal C}, 76(4):222, 2016.

\bibitem{chen2008hawking}
Songbai Chen, Bin Wang, and Rukeng Su.
\newblock Hawking radiation in a d-dimensional static spherically symmetric
  black hole surrounded by quintessence.
\newblock {\em Physical Review D}, 77(12):124011, 2008.

\bibitem{azreg2013thermodynamical}
Mustapha Azreg-A{\"\i}nou and Manuel~E Rodrigues.
\newblock Thermodynamical, geometrical and poincar{\'e} methods for charged
  black holes in presence of quintessence.
\newblock {\em Journal of High Energy Physics}, 2013(9):146, 2013.

\bibitem{chakrabarty2019scalar}
Hrishikesh Chakrabarty, Ahmadjon Abdujabbarov, and Cosimo Bambi.
\newblock Scalar perturbations and quasi-normal modes of a nonlinear
  magnetic-charged black hole surrounded by quintessence.
\newblock {\em The European Physical Journal C}, 79(3):1--11, 2019.

\bibitem{kiselev2003quintessential}
VV~Kiselev.
\newblock Quintessential solution of dark matter rotation curves and its
  simulation by extra dimensions.
\newblock {\em arXiv preprint gr-qc/0303031}, 2003.

\bibitem{nam2018non}
Cao~H Nam.
\newblock On non-linear magnetic-charged black hole surrounded by quintessence.
\newblock {\em General Relativity and Gravitation}, 50(6):57, 2018.

\bibitem{li2012galactic}
Ming-Hsun Li and Kwei-Chou Yang.
\newblock Galactic dark matter in the phantom field.
\newblock {\em Physical Review D}, 86(12):123015, 2012.

\bibitem{xu2018kerr}
Zhaoyi Xu, Xian Hou, and Jiancheng Wang.
\newblock Kerr--anti-de sitter/de sitter black hole in perfect fluid dark
  matter background.
\newblock {\em Classical and Quantum Gravity}, 35(11):115003, 2018.

\bibitem{ayon2000bardeen}
Eloy Ay{\'o}n-Beato and Alberto Garc{\i}a.
\newblock The bardeen model as a nonlinear magnetic monopole.
\newblock {\em Physics Letters B}, 493(1-2):149--152, 2000.

\bibitem{newman1965note}
Ezra~T Newman and AI~Janis.
\newblock Note on the kerr spinning-particle metric.
\newblock {\em Journal of Mathematical Physics}, 6(6):915--917, 1965.

\bibitem{toshmatov2017generic}
Bobir Toshmatov, Zden{\v{e}}k Stuchl{\'\i}k, and Bobomurat Ahmedov.
\newblock Generic rotating regular black holes in general relativity coupled to
  nonlinear electrodynamics.
\newblock {\em Physical Review D}, 95(8):084037, 2017.

\bibitem{kumar2018rotating}
Rahul Kumar and Sushant~G Ghosh.
\newblock Rotating black hole in rastall theory.
\newblock {\em The European Physical Journal C}, 78(9):750, 2018.

\bibitem{xu2017kerr}
Zhaoyi Xu and Jiancheng Wang.
\newblock Kerr-newman-ads black hole in quintessential dark energy.
\newblock {\em Physical Review D}, 95(6):064015, 2017.

\bibitem{xu2020black}
Zhaoyi Xu, Xiaobo Gong, and Shuang-Nan Zhang.
\newblock Black hole immersed dark matter halo.
\newblock {\em Physical Review D}, 101(2):024029, 2020.

\bibitem{shaikh2019black}
Rajibul Shaikh.
\newblock Black hole shadow in a general rotating spacetime obtained through
  newman-janis algorithm.
\newblock {\em Physical Review D}, 100(2):024028, 2019.

\bibitem{jusufi2020rotating}
Kimet Jusufi, Mubasher Jamil, Hrishikesh Chakrabarty, Qiang Wu, Cosimo Bambi,
  and Anzhong Wang.
\newblock Rotating regular black holes in conformal massive gravity.
\newblock {\em Physical Review D}, 101(4):044035, 2020.

\bibitem{zhang2020bardeen}
He-Xu Zhang, Yuan Chen, Peng-Zhang He, Qi-Qi Fan, and Jian-Bo Deng.
\newblock Bardeen black hole surrounded by perfect fluid dark matter.
\newblock {\em arXiv preprint arXiv:2007.09408}, 2020.

\bibitem{azreg2014generating}
Mustapha Azreg-A{\"\i}nou.
\newblock Generating rotating regular black hole solutions without
  complexification.
\newblock {\em Physical Review D}, 90(6):064041, 2014.

\bibitem{carter1968global}
Brandon Carter.
\newblock Global structure of the kerr family of gravitational fields.
\newblock {\em Physical Review}, 174(5):1559, 1968.

\bibitem{chandrasekhar1998mathematical}
Subrahmanyan Chandrasekhar.
\newblock {\em The mathematical theory of black holes}, volume~69.
\newblock Oxford University Press, 1998.

\bibitem{hou2018rotating}
Xian Hou, Zhaoyi Xu, and Jiancheng Wang.
\newblock Rotating black hole shadow in perfect fluid dark matter.
\newblock {\em Journal of Cosmology and Astroparticle Physics}, 2018(12):040,
  2018.

\bibitem{hioki2009measurement}
Kenta Hioki and Kei-ichi Maeda.
\newblock Measurement of the kerr spin parameter by observation of a compact
  object’s shadow.
\newblock {\em Physical Review D}, 80(2):024042, 2009.

\bibitem{wei2013observing}
Shao-Wen Wei and Yu-Xiao Liu.
\newblock Observing the shadow of einstein-maxwell-dilaton-axion black hole.
\newblock {\em Journal of Cosmology and Astroparticle Physics}, 2013(11):063,
  2013.

\bibitem{amarilla2012shadow}
Leonardo Amarilla and Ernesto~F Eiroa.
\newblock Shadow of a rotating braneworld black hole.
\newblock {\em Physical Review D}, 85(6):064019, 2012.

\bibitem{kardashev2013radioastron}
NS~Kardashev, VV~Khartov, VV~Abramov, V~Yu Avdeev, AV~Alakoz, Yu~A Aleksandrov,
  S~Ananthakrishnan, VV~Andreyanov, AS~Andrianov, NM~Antonov, et~al.
\newblock “radioastron”-a telescope with a size of 300 000 km: Main
  parameters and first observational results.
\newblock {\em Astronomy Reports}, 57(3):153--194, 2013.

\end{thebibliography}
\end{document}